\documentclass[11pt]{article}

\usepackage{acl}

\usepackage{times}
\usepackage{latexsym}
\usepackage{amsmath}
\DeclareMathOperator{\doop}{do}

\usepackage[T1]{fontenc}
\usepackage[utf8]{inputenc}
\usepackage{microtype}
\usepackage{inconsolata}
\usepackage{graphicx}

\usepackage{float}
\usepackage{placeins}

\usepackage{booktabs}   
\usepackage{multirow}   
\usepackage{adjustbox}  
\usepackage[table]{xcolor} 
\usepackage{ulem}       
\usepackage{stfloats}
\usepackage{subcaption}
\definecolor{baseblue}{HTML}{1F77B4}
\definecolor{aceorange}{HTML}{FF7F0E}
\definecolor{culturegreen}{HTML}{2CA02C}
\definecolor{causalred}{HTML}{D62728}
\definecolor{barbasegray}{HTML}{8FA1B2}
\definecolor{baraligned}{HTML}{2A9D8F}
\definecolor{barflipped}{HTML}{D62828}
\definecolor{barstereotyping}{HTML}{F77F00}
\definecolor{barerasure}{HTML}{003049}

\title{ACE-Align: Attribute Causal Effect Alignment for Cultural Values under Varying Persona Granularities}

\author{
  \textbf{Jiatang Luo}\textsuperscript{1,2,3},
  \textbf{Bingbing Xu}\textsuperscript{2,3}\thanks{~~Corresponding author.},
  \textbf{Rongxin Chen}\textsuperscript{2,3},
  \textbf{Xiaoyan Zhao}\textsuperscript{4,5},
  \\
  \textbf{Yang Zhang}\textsuperscript{4},
  \textbf{Liang Pang}\textsuperscript{2,3},
  \textbf{Zhiyong Huang}\textsuperscript{4},
  \textbf{Huawei Shen}\textsuperscript{2,3},
  \\
  \textsuperscript{1}School of Advanced Interdisciplinary Sciences, University of Chinese Academy of Sciences, China \\
  \textsuperscript{2}State Key Laboratory of AI Safety, Institute of Computing Technology, CAS \\
  \textsuperscript{3}University of Chinese Academy of Sciences \\
  \textsuperscript{4}National University of Singapore \\
  \textsuperscript{5}The Chinese University of Hong Kong \\
  \texttt{luojiatang23@mails.ucas.ac.cn,\{xubingbing,chenrongxin24s,pangliang\}@ict.ac.cn}
}

\begin{document}
\maketitle

\begin{abstract}
Ensuring that large language models (LLMs) reflect diverse cultural values is important for globally deployed NLP systems.
However, existing approaches often treat cultural groups as homogeneous and overlook within-group heterogeneity arising from intersecting demographic attributes, leading to unstable behavior under varying persona granularity.
To address this gap, we propose \textbf{ACE-Align} (\textbf{A}ttribute \textbf{C}ausal \textbf{E}ffect Alignment), a causally inspired framework based on controlled persona edits that aligns how specific demographic attributes shift different cultural values, rather than treating each culture as a homogeneous group.
We evaluate ACE-Align across 14 countries spanning five continents, with personas specified by subsets of four attributes (gender, education, residence, and marital status) and granularity instantiated by the number of specified attributes.
Across all persona granularities, ACE-Align consistently outperforms baselines.
Moreover, in within-survey comparisons, it reduces the average Global North--South alignment gap from 3.40 to 1.11 points on WVS and from 2.53 to 0.85 points on ISSP. Code and dataset are released at this \href{https://github.com/Wells-Luo/ACE-Align}{link}.
\end{abstract}

\section{Introduction}

As large language models (LLMs) become globally deployed, aligning their outputs with diverse cultural values has become an important challenge for socially responsible NLP systems \cite{cao2023assessing}.
Despite substantial progress in instruction tuning and preference optimization, a persistent alignment deficit remains: a model may appear aligned with a broad cultural or national group, yet still misrepresent sub-populations within that group that are defined by intersecting demographic attributes \cite{santurkar2023whose}.
This deficit is especially consequential in regions such as Africa and Southeast Asia, where uneven digital access and scarce high-quality text contribute to weaker representation in the corpora used to train LLMs, increasing the risk of cultural erasure and stereotyping \cite{decoupes2025evaluation}.

\begin{figure}[t]
    \centering
    \includegraphics[width=1.0\linewidth]
    {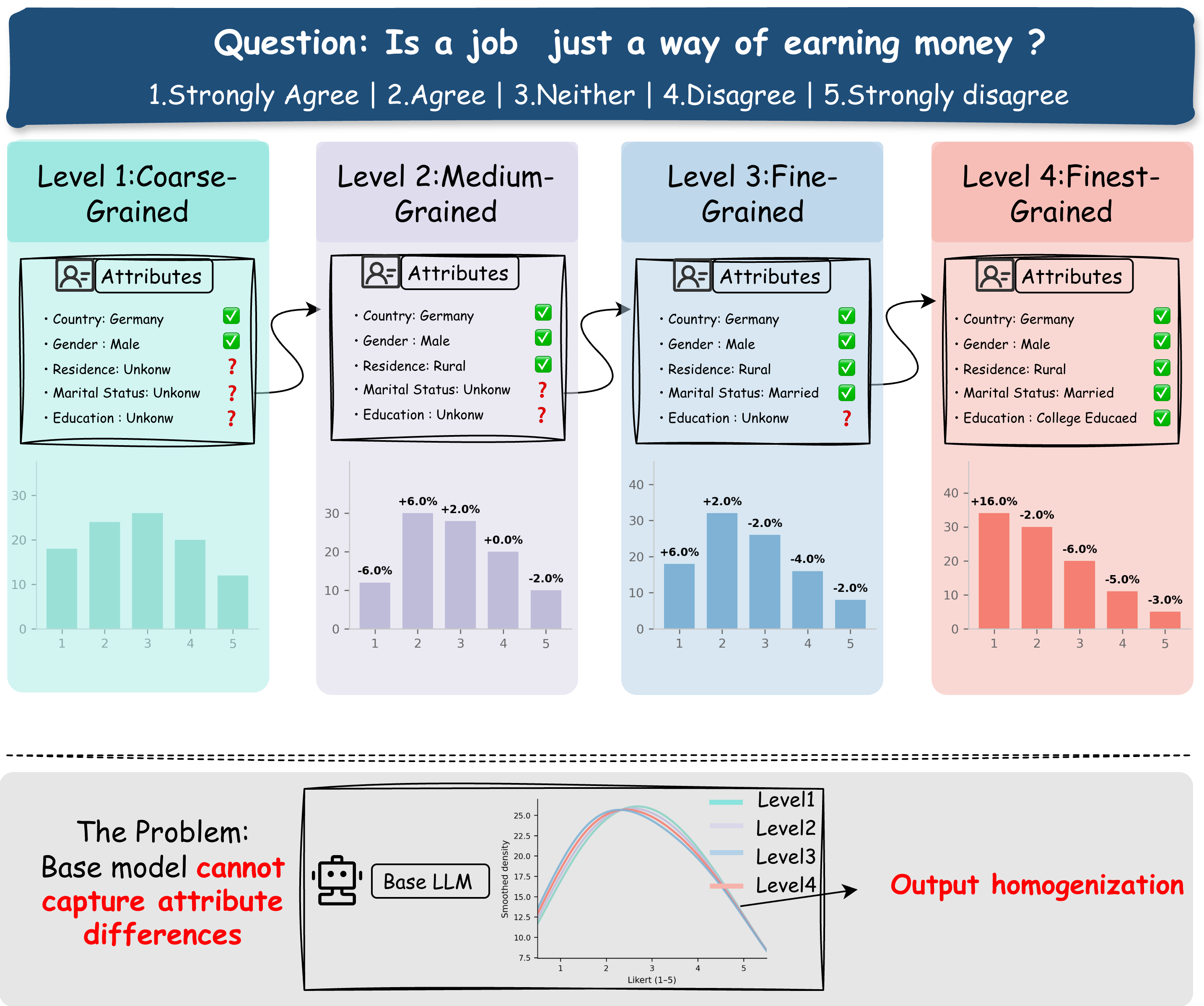}
    \caption{Within a single culture, personas at different granularities exhibit heterogeneous values, which current LLMs fail to consistently capture.}
    \label{fig:intro_res}
\end{figure}

This challenge becomes more pronounced when persona conditioning varies in granularity.
As illustrated in Figure~\ref{fig:intro_res}, personas may describe broad sub-populations using country and a single demographic attribute, or progressively narrower sub-populations by combining multiple intersecting attributes \cite{chen2026multi}.
A model may behave plausibly for a coarse-grained persona (e.g., \textit{German man}), yet degrade or become inconsistent when conditioned on a finer-grained persona that adds residence, education, and marital status (e.g., \textit{a rural, college-educated, married German man}).
Consistent with this challenge, Table~\ref{tab:main_res} shows that all evaluated methods obtain lower average alignment scores at $G{=}4$ than at $G{=}1$.
This suggests that cultural alignment methods should remain reliable across persona granularities, rather than being optimized for a single granularity level.

Existing alignment strategies often fit a direct mapping from persona descriptions to target responses, without explicitly constraining how responses should shift when persona attributes are combined \cite{xu2025self,alkhamissi2024investigating}.
This can lead to unstable attribute interactions: models may exhibit \textit{stereotyping} (over-amplifying demographic influences)~\cite{anthis2025llm}, \textit{erasure} (understating or homogenizing demographic differences)~\cite{santurkar2023whose}, or \textit{flipped signs} (reversing the direction of demographic influence)~\cite{talat2022machine}.

To address these failures, we argue that alignment across persona granularities requires explicitly modeling how demographic attributes are associated with shifts in cultural values.
Inspired by counterfactual reasoning in causal inference, we operationalize this with controlled persona edits: one attribute is toggled while the country, question, and other specified attributes are held fixed.
These contrasts provide survey-grounded estimates of how each attribute is associated with response shifts, but they should not be interpreted as randomized real-world intervention effects.

In this paper, we instantiate this perspective with \textbf{ACE-Align} (\textbf{A}ttribute \textbf{C}ausal \textbf{E}ffect Alignment), which trains LLMs to match model-side controlled persona-edit effects with survey-grounded attribute effects.
Since effect matching only constrains relative response shifts, it can admit multiple absolute prediction distributions with the same attribute effect.
ACE-Align therefore adds a lightweight anchoring loss that grounds absolute predictions while preserving the learned response shifts.
This encourages the model to match both the direction and magnitude of attribute influence while remaining calibrated to survey responses.

Our contributions are summarized as follows:
\begin{itemize}
    \item \textbf{Investigating Persona Granularity.} We formalize the challenge of cultural alignment across varying persona granularities and reveal the performance instability of LLMs under intersecting demographic attributes.
    \item \textbf{ACE-Align Framework.} We propose \textbf{ACE-Align}, which matches controlled persona-edit effects to survey-grounded attribute effects while anchoring absolute predictions, improving consistency across persona granularities.
    \item \textbf{Global Equity and Bias Diagnosis.} Following the Brandt-line North--South division~\citep{brandt1980north}, we compare comparatively industrialized economies (Global North) with developing and emerging economies (Global South). Within WVS and ISSP separately, ACE-Align narrows the resulting alignment gap and reduces attribute-effect errors such as stereotyping and cultural erasure.
\end{itemize}

\begin{figure*}[t]
    \centering
    \includegraphics[width=\textwidth]{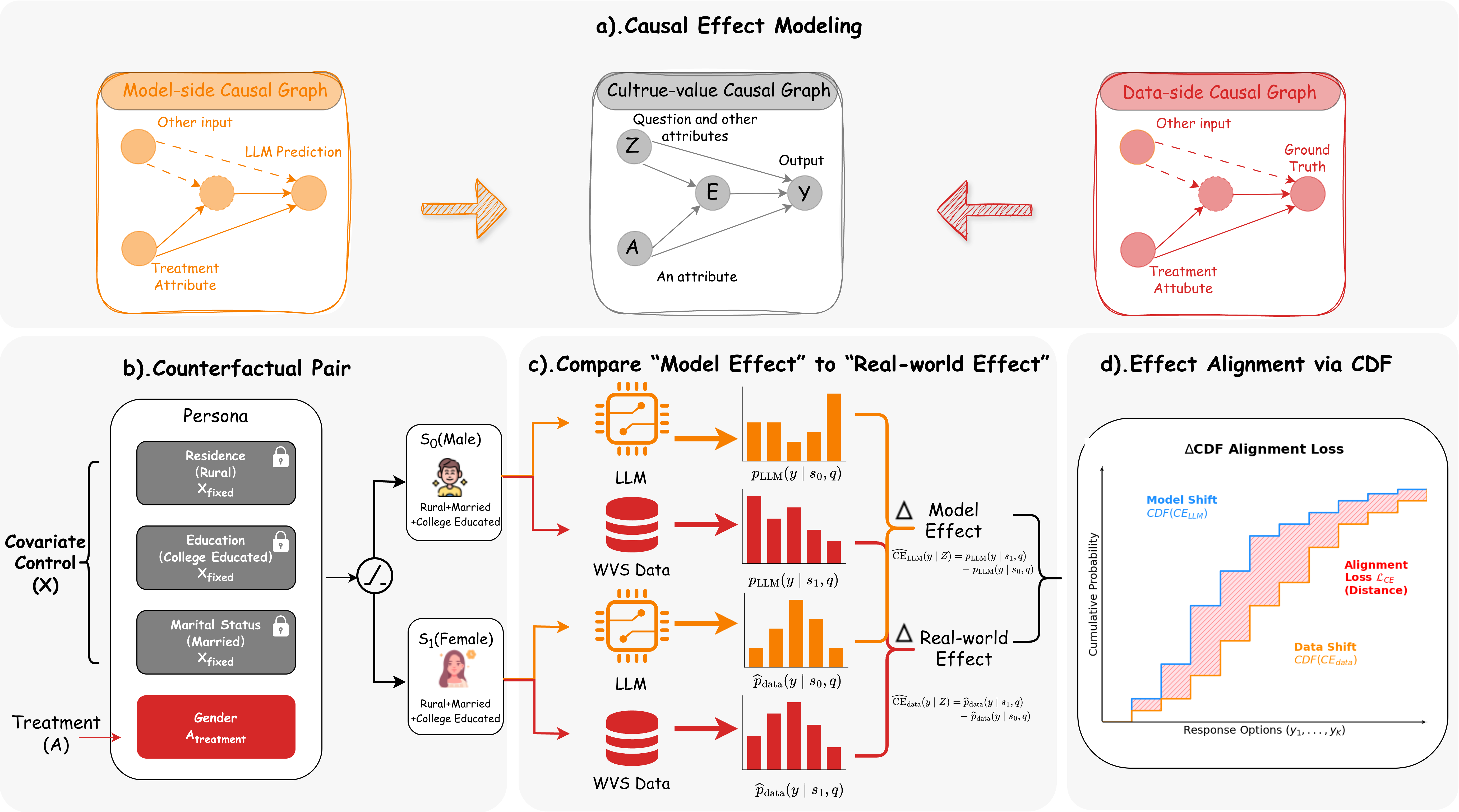}
    \caption{\textbf{The ACE-Align framework.} We optimize cultural alignment at the finest granularity ($G=4$) to keep the remaining persona covariates fixed when constructing attribute-effect contrasts. By aligning the model's controlled response shifts ($\mathrm{CE}_{\mathrm{LLM}}$) with survey-grounded attribute effects ($\mathrm{CE}_{\mathrm{data}}$) via $\Delta$CDF distance, the model learns response-shift patterns that maintain consistency across persona granularities.}
    \label{fig:ace_align_framework}
\end{figure*}

\section{Related Work}

\subsection{Cultural Alignment}
Recent work evaluates and improves the cultural alignment of LLMs using survey-style benchmarks and cross-cultural opinion distributions \citep{cao2023assessing,santurkar2023whose,wang2025sociobench,sukiennik2025evaluation,kovavc2023large,tao2024cultural}.
Existing approaches span prompting and inference-time strategies \citep{alkhamissi2024investigating} and alignment training via supervised fine-tuning on survey data \citep{li2024culturellm,masoud2025cultural,du2025simvbg}.
These methods often target culture at a fixed resolution, which can mask within culture heterogeneity when persona attributes are specified at different granularity levels \citep{decoupes2025evaluation}.
Our work focuses on consistency across persona granularities and treats demographic attributes as compositional factors that drive heterogeneous cultural topic preferences.

\subsection{Causal Perspectives for Debiasing}
Causal formulations have been used to analyze and mitigate bias by comparing interventions and counterfactual contrasts rather than relying only on correlations~\cite{chaudhary2025certifying}.
Prior work has applied this perspective to dataset bias detection via causal invariance and active selection \citep{sun2024causal}, reasoning debiasing through counterfactual chains \citep{wu2024decot}, and annotation artifact mitigation through causal mediation analysis and targeted unlearning \citep{lim2024identifying}.
Other studies localize where social bias is mediated inside language models \citep{vig2020investigating}, or use front-door adjustment for prompting-time interventions when model parameters are inaccessible \citep{zhang2025causal}.
Our work differs by applying controlled persona edits to cultural alignment under persona conditioning, aligning attribute-level response shifts to improve consistency across persona granularities.

\section{Method}

\subsection{Task Formulation}
\label{sec:task}

We investigate cultural alignment of model responses to culture topics under persona conditioning at varying granularity.
Here, a persona denotes a subgroup profile defined by a set of attributes, rather than a specific individual identity.
We instantiate persona granularity by the number of specified attributes in the profile.

Let $\mathcal{A}=\{a_1,\dots,a_M\}$ denote the universe of available attributes. A persona $g$ is an instantiated subset of $\mathcal{A}$, where selected attributes are assigned specific values. The persona granularity is defined as the number of specified attributes, denoted by the scalar $G = |g|$.

Given a value-laden question $q$, the response options form an ordered discrete set $\mathcal{Y}=\{1,\dots,K\}$. The model $\theta$ outputs a categorical distribution over these options:
\begin{equation}
p_\theta(y \mid g, q), \quad y\in\mathcal{Y},
\end{equation}
where $\theta$ represents the model parameters.

\paragraph{Obtaining option probabilities from the LLM.}
To instantiate the categorical distribution over the discrete option set $\mathcal{Y}$, we compute the likelihood of each option under the LLM and renormalize over $\mathcal{Y}$. Specifically, for each option $y\in\mathcal{Y}$, let $\mathbf{t}(y)=(t_1,\ldots,t_{m_y})$ be the token sequence of its canonical verbalization under the model tokenizer. Given the formatted $\text{prompt}(g,q)$, we define the option log-likelihood as
\begin{equation}
\ell_\theta(y \mid g,q)=\sum_{j=1}^{m_y}\log p_\theta\!\left(t_j \mid \text{prompt}(g,q), t_{<j}\right),
\label{eq:option_ll}
\end{equation}
and obtain a categorical distribution by renormalizing within $\mathcal{Y}$:
\begin{equation}
p_\theta(y \mid g,q)=\frac{\exp(\ell_\theta(y \mid g,q))}{\sum_{y'\in\mathcal{Y}}\exp(\ell_\theta(y' \mid g,q))}.
\label{eq:option_prob}
\end{equation}

\subsection{ACE-Align Framework}
\label{sec:ace_align}

To capture how demographic attributes influence different cultural topics, we propose to align attribute causal effects.
By learning survey-grounded response shifts associated with each attribute, the model can compositionally generalize to novel attribute combinations.
As illustrated in Figure~\ref{fig:ace_align_framework}, we introduce \textbf{ACE-Align} (\textbf{A}ttribute \textbf{C}ausal \textbf{E}ffect Alignment), a framework that explicitly estimates and aligns these effects during training.
In training, we instantiate contexts using the finest granularity level ($G=4$), so that when toggling a treatment attribute $A$, all remaining demographic attributes are held fixed in $X$.
This provides the strongest available covariate control in our persona design and reduces variation from omitted specified attributes, yielding more reliable shifts associated with the toggled attribute.

\subsubsection{Causal Effect Modeling}

\paragraph{Causal Graph.}
For a culture topic question $q$ with ordinal options $\mathcal{Y}$, we treat the selected option as a discrete outcome variable $Y \in \mathcal{Y}$.
We choose one attribute $A \in \mathcal{A}$ as the treatment and denote the remaining persona specification as $X$.
We define the context as $Z=(X,q)$, where $X$ contains all other specified attributes and $q$ specifies the culture topic question.
Figure~\ref{fig:ace_align_framework}a shows our directed acyclic graph (DAG) \cite{pearl2009causality}, where $A$ and $Z$ directly influence $Y$.
To account for complex and often unobservable dependencies between demographics and cultural contexts, we include a latent factor $E$ that represents interactions between $A$ and $Z$~\cite{zhao2026nextquill}.
We operationalize the attribute effect by toggling only $A$ while holding $Z$ fixed.

\paragraph{Identification Assumption.}
For each treated attribute $A$, we view controlled persona editing as an operational intervention on the persona specification: $\doop(A=a)$ sets the value of one persona attribute while keeping $Z=(X,q)$ fixed.
In the survey data, we approximate this controlled-edit distribution by the corresponding conditional distribution:
\begin{equation}
p_a^{\doop}(y\mid Z)\approx p(Y=y\mid A=a,Z).
\label{eq:identification}
\end{equation}

Importantly, our causal interpretation is defined in the persona-conditioning space: on the model side, changing $A$ while holding $Z$ fixed is a controlled input intervention, whereas on the survey side, because demographic attributes are not randomized, we use matched conditional response distributions as survey-grounded approximations of the corresponding controlled-edit effects.
This approximation supports controlled persona-edit contrasts, but should not be interpreted as identifying randomized real-world causal effects.

\paragraph{Binary treatment coding.}
For each treatment attribute, we define a binary variable $A\in\{0,1\}$ and estimate attribute effects via controlled persona edits.
We consistently define the causal effect direction as
$p(Y\mid A{=}1,Z) - p(Y\mid A{=}0,Z)$.
The concrete attribute-value coding is reported in Appendix~\ref{sec:attribute_coding}.

\subsubsection{Attribute Causal Effect}
For a target attribute $A$ under a fixed context $Z=(X,q)$, let $r\in\{m,d\}$ index the model and survey data, with $p_m=p_\theta$ and $p_d=p_{\text{data}}$.
For compactness, write $p_r^a(y\mid Z)=p_r(Y=y\mid\doop(A=a),Z)$.
For each answer option $y\in\mathcal{Y}$ and $r\in\{m,d\}$, the option-level causal effect is:
\begin{equation}
\mathrm{CE}_{r}(y\mid A,Z)
=
p_r^1(y\mid Z)-p_r^0(y\mid Z).
\label{eq:ce_definition}
\end{equation}
Under Eq.~\eqref{eq:identification}, we operationalize this effect with two toggled persona specifications $s_1=(A=1,X)$ and $s_0=(A=0,X)$:
\begin{equation}
\widehat{\mathrm{CE}}_{r}(y\mid A,Z)
=
\widehat{p}_{r}^{\,1}(y\mid Z)
-
\widehat{p}_{r}^{\,0}(y\mid Z).
\label{eq:empirical_ce}
\end{equation}
Here $\widehat{p}_{r}^{\,a}(y\mid Z)=\widehat{p}_{r}(y\mid s_a,q)$. For the model, $\widehat{p}_{m}$ is obtained from a forward pass; for survey data, $\widehat{p}_{d}$ is the empirical response distribution of matched respondents.

\subsubsection{Effect Alignment Objective}

\paragraph{Aligning Attribute Causal Effects.}
The core of our framework is to ensure the model mimics how human responses \textit{shift} when a specific attribute $A$ is toggled. This is achieved via three steps:

\begin{itemize}
    \item \textbf{Step 1: Vector Representation.} For each $(A,Z)$ and $r\in\{m,d\}$, we stack the option-level effects from Eq.~\eqref{eq:empirical_ce} into a vector over $k=1,\ldots,K$:
    \begin{equation}
    \widehat{c}_{r,k}(A,Z)
    =
    \widehat{\mathrm{CE}}_{r}(k\mid A,Z).
    \label{eq:ce_vector}
    \end{equation}

    \item \textbf{Step 2: Cumulative Transformation.} Since the response options $\mathcal{Y}$ are ordinal, we cumulatively sum the option-level effects to capture distributional shifts along the ordinal scale:
    \begin{equation}
    \Delta F_{r}(k\mid A,Z)
    =
    \sum_{i=1}^{k}
    \widehat{c}_{r,i}(A,Z).
    \label{eq:cdf_shift}
    \end{equation}

    \item \textbf{Step 3: Measuring Alignment.} We measure the distance between model and data shifts across cumulative levels:
    \begin{equation}
    \begin{aligned}
    d_{\text{CDF}}(A,Z)
    &=
    \frac{1}{K}
    \sum_{k=1}^{K}
    \bigl|
    \Delta F_{m}(k\mid A,Z)\\
    &\quad
    -
    \Delta F_{d}(k\mid A,Z)
    \bigr|.
    \end{aligned}
    \label{eq:cdf_distance}
    \end{equation}

\end{itemize}

The total attribute-effect matching loss $\mathcal{L}_{\text{CE}}$ is defined as:

\begin{equation}
\mathcal{L}_{\text{CE}} = \frac{1}{|\mathcal{D}|} \sum_{(A,Z)\in\mathcal{D}} d_{\text{CDF}}(A,Z).
\label{eq:ace_loss}
\end{equation}

Here $\mathcal{D}$ denotes the set of valid $(A,Z)$ pairs, where $A$ is a demographic attribute and $Z=(X,q)$ is a context.
We include $(A,Z)$ only when both toggled personas $s_1=(A{=}1,X)$ and $s_0=(A{=}0,X)$ are supported by sufficient survey respondents for question $q$. Concretely, we discard a context if either subgroup has fewer than 10 matched respondents.
In training, we construct $Z=(X,q)$ using the finest-grained personas ($G=4$), so that when toggling $A$ all remaining demographic attributes are held fixed in $X$, matching the controlled edit construction described above.

\paragraph{Anchoring the Model Response.}
The attribute-effect matching loss $\mathcal{L}_{\text{CE}}$ constrains the relative shift between a toggled persona pair under a fixed context, but is under-determined when optimized alone, as multiple absolute distributions can realize the same shift.
To provide absolute reference points for model predictions, we introduce a per-persona anchoring loss:
\begin{equation}
\ell_{\text{anchor}}(s,q) = -\log p_\theta(y^*_{s,q} \mid s, q),
\end{equation}
where $y^*_{s,q}$ denotes the empirical mode of the survey responses under persona $s$ and question $q$.
For each $(A,Z)\in\mathcal{D}$, this loss is applied to both endpoints $s_1=(A{=}1,X)$ and $s_0=(A{=}0,X)$ of the controlled persona edit.

\paragraph{Overall Objective.}
The final training objective integrates absolute response grounding with attribute-effect matching:
\begin{equation}
\mathcal{L} = \alpha\,\mathcal{L}_{\text{anchor}} + \beta\,\mathcal{L}_{\text{CE}},
\label{eq:overall_obj}
\end{equation}
where $\alpha$ and $\beta$ balance absolute response grounding and attribute-effect matching. We define $\mathcal{L}_{\text{anchor}}$ by aggregating $\ell_{\text{anchor}}(s,q)$ over both endpoints $s_1$ and $s_0$ for each $(A,Z)\in\mathcal{D}$.

\section{Experiments}
In this section, we report comprehensive experimental results on cultural alignment under persona conditioning at varying persona granularities. Our analysis is guided by the following three research questions \textbf{(RQs)}:

\begin{itemize}
    \item \textbf{RQ1: Does ACE-Align consistently improve cultural alignment across varying persona granularities?}
    \item \textbf{RQ2: How robust is ACE-Align under low-resource conditions?}
    \item \textbf{RQ3: What mechanism drives the effectiveness of ACE-Align?}
\end{itemize}

\begin{table*}[!t]
\centering
\small
\setlength{\tabcolsep}{3pt}
\renewcommand{\arraystretch}{1.12}
\resizebox{\textwidth}{!}{
\begin{tabular}{l|ccc|ccc|ccc|ccc|cc|l}
\toprule
{} & \multicolumn{3}{c|}{\textbf{America}} & \multicolumn{3}{c|}{\textbf{Europe}} & \multicolumn{3}{c|}{\textbf{Asia}} & \multicolumn{3}{c|}{\textbf{Africa}} & \multicolumn{2}{c|}{\textbf{Oceania}} & \multicolumn{1}{c}{\textbf{Avg}} \\
\cline{2-15}
{} & \textbf{CHL} & \textbf{MEX} & \textbf{USA} & \textbf{DEU} & \textbf{GBR} & \textbf{RUS} & \textbf{IND} & \textbf{JPN} & \textbf{PHL} & \textbf{EGY} & \textbf{ETH} & \textbf{NGA} & \textbf{AUS} & \textbf{NZL} & {}\\
\midrule

\multicolumn{16}{c}{\emph{\textbf{$G=1$}}} \\
\midrule
\textbf{Base model} & 82.16 & 82.75 & 84.04 & 83.74 & 84.71 & 80.46 & 81.37 & 82.89 & 80.70 & 72.87 & 74.74 & 77.35 & 84.44 & 85.30 & 81.25 \\
\textbf{AdPrompt} & 83.20 & 83.37 & 84.37 & 83.82 & 84.47 & 81.47 & 82.36 & 83.03 & 81.53 & 73.05 & 74.93 & 78.36 & 84.15 & 84.86 & 81.64 \scriptsize{(+0.39)} \\
\textbf{DPO} & 81.70 & 82.75 & 83.16 & 83.69 & 84.25 & 81.01 & 81.56 & 82.90 & 81.42 & 74.35 & 76.53 & 79.20 & 84.03 & 85.29 & 81.56 \scriptsize{(+0.31)} \\
\textbf{PPO} & 81.27 & 81.22 & 82.61 & 82.78 & 83.32 & 81.20 & 80.57 & 82.04 & 81.55 & 72.10 & 74.03 & 76.11 & 82.94 & 84.19 & 80.42 \scriptsize{(-0.83)} \\
\textbf{GRPO} & 82.35 & 80.38 & 84.92 & 83.92 & 85.92 & 80.56 & 84.73 & 82.90 & 83.28 & \underline{80.88} & \textbf{85.49} & \textbf{86.12} & 86.10 & 86.31 & 83.85 \scriptsize{(+2.60)} \\
\textbf{SimLLM} & 83.55 & \underline{84.63} & \underline{86.21} & 85.49 & \underline{86.43} & \underline{83.12} & 83.20 & \underline{84.52} & 83.25 & 80.11 & \underline{83.48} & \underline{85.19} & \underline{86.48} & \textbf{87.17} & \underline{84.49} \scriptsize{(+3.24)} \\
\textbf{CultureSPA} & 81.72 & 83.68 & 83.97 & 82.44 & 83.20 & 81.67 & 83.23 & 82.25 & 82.04 & 73.48 & 77.50 & 81.13 & 82.95 & 83.26 & 81.61 \scriptsize{(+0.36)} \\
\textbf{Anchor-only} & \underline{84.18} & 79.58 & 81.63 & 81.91 & 80.03 & 79.82 & 80.60 & 81.77 & 82.49 & 79.58 & 73.56 & 74.62 & 82.12 & 82.30 & 80.30 \scriptsize{(-0.95)} \\
\textbf{Causal-only} & 83.65 & 82.86 & \underline{86.21} & \underline{85.53} & 84.84 & 82.28 & \underline{86.37} & 81.69 & \underline{84.50} & 72.81 & 76.68 & 79.56 & 84.65 & 85.92 & 82.68 \scriptsize{(+1.43)} \\
\rowcolor{gray!10} \textbf{ACE-Align} & \textbf{85.17} & \textbf{86.39} & \textbf{87.75} & \textbf{87.54} & \textbf{87.23} & \textbf{84.53} & \textbf{87.57} & \textbf{85.19} & \textbf{87.24} & \textbf{85.00} & 79.33 & 84.25 & \textbf{87.98} & \underline{86.99} & \textbf{85.87} \scriptsize{(+4.62)} \\

\midrule
\multicolumn{16}{c}{\emph{\textbf{$G=2$}}} \\
\midrule
\textbf{Base model} & 82.24 & 82.94 & 84.31 & 83.81 & 84.69 & 81.05 & 81.19 & 83.07 & 81.11 & 72.92 & 74.12 & 76.95 & 84.39 & 85.24 & 81.29 \\
\textbf{AdPrompt} & 83.10 & 83.22 & 84.24 & 83.80 & 84.38 & 81.60 & 82.00 & 83.08 & 81.52 & 73.24 & 74.46 & 78.00 & 84.08 & 84.76 & 81.53 \scriptsize{(+0.24)} \\
\textbf{DPO} & 81.76 & 82.81 & 83.59 & 83.81 & 84.32 & 81.60 & 81.38 & 83.03 & 81.82 & 74.69 & 75.82 & 78.45 & 83.94 & 85.26 & 81.59 \scriptsize{(+0.30)} \\
\textbf{PPO} & 81.20 & 80.89 & 82.37 & 82.62 & 83.11 & 81.35 & 80.19 & 81.66 & 81.60 & 72.20 & 73.69 & 76.13 & 82.60 & 83.80 & 80.24 \scriptsize{(-1.05)} \\
\textbf{GRPO} & 81.90 & 79.54 & 84.83 & 83.65 & 85.78 & 80.45 & 84.21 & 82.83 & 83.23 & \underline{81.16} & \textbf{85.15} & \textbf{85.84} & 85.90 & 86.09 & 83.61 \scriptsize{(+2.32)} \\
\textbf{SimLLM} & 83.51 & \underline{84.59} & \underline{86.23} & 85.44 & \underline{86.28} & \underline{83.45} & 82.90 & \underline{84.61} & 83.57 & 79.93 & \underline{82.93} & 84.70 & \underline{86.25} & \textbf{86.95} & \underline{84.38} \scriptsize{(+3.09)} \\
\textbf{CultureSPA} & 81.58 & 83.26 & 83.64 & 82.21 & 82.87 & 81.62 & 82.82 & 82.07 & 82.10 & 73.58 & 77.30 & 80.68 & 82.70 & 83.00 & 81.39 \scriptsize{(+0.10)} \\
\textbf{Anchor-only} & 83.28 & 79.19 & 81.89 & 81.73 & 80.24 & 79.92 & 79.87 & 81.24 & 82.26 & 79.52 & 72.85 & 74.15 & 81.91 & 82.16 & 80.02 \scriptsize{(-1.27)} \\
\textbf{Causal-only} & \underline{83.78} & 82.79 & 86.15 & \underline{85.60} & 85.00 & 82.59 & \underline{86.35} & 81.75 & \underline{84.98} & 73.08 & 76.27 & 79.31 & 84.77 & 86.15 & 82.76 \scriptsize{(+1.47)} \\
\rowcolor{gray!10} \textbf{ACE-Align} & \textbf{85.03} & \textbf{86.05} & \textbf{87.60} & \textbf{87.47} & \textbf{87.13} & \textbf{84.79} & \textbf{87.19} & \textbf{85.02} & \textbf{87.41} & \textbf{85.19} & 79.19 & \underline{84.74} & \textbf{87.68} & \underline{86.73} & \textbf{85.80} \scriptsize{(+4.51)} \\

\midrule
\multicolumn{16}{c}{\emph{\textbf{$G=3$}}} \\
\midrule
\textbf{Base model} & 82.03 & 82.70 & 84.18 & 83.59 & 84.40 & 80.88 & 80.67 & 82.80 & 81.09 & 72.84 & 73.54 & 76.39 & 84.02 & 84.79 & 80.99 \\
\textbf{AdPrompt} & 82.82 & 82.87 & 83.84 & 83.59 & 83.99 & 81.42 & 81.42 & 82.90 & 81.26 & 73.27 & 73.84 & 77.36 & 83.69 & 84.32 & 81.19 \scriptsize{(+0.20)} \\
\textbf{DPO} & 81.73 & 82.73 & 83.75 & 83.83 & 84.26 & 81.51 & 81.02 & 82.91 & 82.02 & 74.74 & 75.28 & 77.98 & 83.80 & 85.05 & 81.47 \scriptsize{(+0.48)} \\
\textbf{PPO} & 81.27 & 80.90 & 82.27 & 82.67 & 83.00 & 81.25 & 80.06 & 81.47 & 81.62 & 72.33 & 73.74 & 76.23 & 82.48 & 83.53 & 80.20 \scriptsize{(-0.79)} \\
\textbf{GRPO} & 81.50 & 79.64 & 84.41 & 83.21 & 85.49 & 80.17 & 83.56 & 82.40 & 82.78 & \underline{81.39} & \textbf{84.48} & \textbf{85.50} & 85.40 & 85.49 & 83.24 \scriptsize{(+2.25)} \\
\textbf{SimLLM} & 83.28 & \underline{84.43} & \underline{85.97} & 85.33 & \underline{85.98} & \underline{83.32} & 82.44 & \underline{84.37} & 83.53 & 79.65 & \underline{82.51} & 84.26 & \underline{85.86} & \textbf{86.52} & \underline{84.10} \scriptsize{(+3.11)} \\
\textbf{CultureSPA} & 81.28 & 82.90 & 83.16 & 81.91 & 82.44 & 81.31 & 82.17 & 81.77 & 81.88 & 73.61 & 77.10 & 80.27 & 82.27 & 82.55 & 81.04 \scriptsize{(+0.05)} \\
\textbf{Anchor-only} & 82.23 & 78.39 & 81.82 & 81.52 & 80.57 & 79.65 & 79.27 & 80.56 & 81.92 & 79.15 & 72.82 & 74.14 & 81.71 & 82.12 & 79.71 \scriptsize{(-1.28)} \\
\textbf{Causal-only} & \underline{83.79} & 82.54 & 85.93 & \underline{85.60} & 84.96 & 82.60 & \underline{85.91} & 81.61 & \underline{85.04} & 73.23 & 75.99 & 79.01 & 84.56 & 85.91 & 82.62 \scriptsize{(+1.63)} \\
\rowcolor{gray!10} \textbf{ACE-Align} & \textbf{84.95} & \textbf{85.91} & \textbf{87.34} & \textbf{87.37} & \textbf{87.05} & \textbf{84.73} & \textbf{86.69} & \textbf{84.89} & \textbf{87.30} & \textbf{85.03} & 79.05 & \underline{84.95} & \textbf{87.35} & \underline{86.50} & \textbf{85.65} \scriptsize{(+4.66)} \\

\midrule
\multicolumn{16}{c}{\emph{\textbf{$G=4$}}} \\
\midrule
\textbf{Base model} & 81.82 & 82.64 & 83.83 & 83.32 & 83.93 & 80.61 & 79.94 & 82.54 & 80.87 & 72.67 & 72.80 & 75.72 & 83.52 & 84.15 & 80.60 \\
\textbf{AdPrompt} & 82.57 & 82.72 & 83.38 & 83.26 & 83.49 & 81.13 & 80.65 & 82.70 & 80.89 & 73.23 & 73.13 & 76.57 & 83.18 & 83.72 & 80.76 \scriptsize{(+0.16)} \\
\textbf{DPO} & 81.78 & 83.02 & 83.73 & 83.79 & 84.02 & 81.38 & 80.53 & 82.80 & 82.07 & 74.86 & 74.62 & 77.42 & 83.51 & 84.65 & 81.30 \scriptsize{(+0.70)} \\
\textbf{PPO} & 81.39 & 81.23 & 82.18 & 82.72 & 82.77 & 81.21 & 79.84 & 81.30 & 81.58 & 72.37 & 73.64 & 76.28 & 82.28 & 83.34 & 80.15 \scriptsize{(-0.45)} \\
\textbf{GRPO} & 81.04 & 80.01 & 83.94 & 82.65 & 85.01 & 79.69 & 82.72 & 82.23 & 82.16 & \underline{81.49} & \textbf{84.01} & \textbf{84.80} & 84.78 & 84.77 & 82.81 \scriptsize{(+2.21)} \\
\textbf{SimLLM} & 83.13 & \underline{84.57} & 85.64 & 85.18 & \underline{85.57} & \underline{83.10} & 81.81 & \underline{84.14} & 83.28 & 79.54 & \underline{81.96} & 83.71 & \underline{85.42} & \underline{85.96} & \underline{83.79} \scriptsize{(+3.19)} \\
\textbf{CultureSPA} & 80.93 & 82.88 & 82.72 & 81.46 & 81.89 & 80.90 & 81.18 & 81.44 & 81.38 & 73.67 & 76.73 & 79.64 & 81.66 & 81.81 & 80.59 \scriptsize{(-0.01)} \\
\textbf{Anchor-only} & 81.30 & 77.47 & 81.34 & 81.27 & 80.83 & 79.25 & 78.58 & 80.13 & 81.63 & 78.31 & 72.42 & 74.27 & 81.55 & 82.30 & 79.33 \scriptsize{(-1.27)} \\
\textbf{Causal-only} & \underline{83.76} & 82.43 & \underline{85.75} & \underline{85.41} & 84.70 & 82.45 & \underline{85.11} & 81.48 & \underline{84.86} & 73.35 & 75.60 & 78.53 & 84.14 & 85.40 & 82.36 \scriptsize{(+1.76)} \\
\rowcolor{gray!10} \textbf{ACE-Align} & \textbf{84.90} & \textbf{86.17} & \textbf{87.21} & \textbf{87.16} & \textbf{86.85} & \textbf{84.55} & \textbf{86.10} & \textbf{84.81} & \textbf{87.00} & \textbf{84.67} & 78.53 & \underline{84.77} & \textbf{86.91} & \textbf{86.27} & \textbf{85.42} \scriptsize{(+4.82)} \\

\bottomrule
\end{tabular}
}
\caption{Cultural alignment scores under varying persona granularities $G\in\{1,2,3,4\}$. Best results are shown in bold and second-best results are underlined.}
\label{tab:main_res}
\end{table*}

\subsection{Settings}
\label{sec:settings}

\paragraph{Datasets.}
Following prior work~\cite{cao2023assessing,masoud2025cultural,zhao2024worldvaluesbench,sukiennik2025evaluation}, we assess LLM cultural alignment by subjecting models to standardized sociological surveys under diverse demographic contexts.
We use the \textit{World Values Survey} (WVS) Wave~7 as training data~\cite{haerpfer2022world}.
For evaluation, we test on ISSP~\cite{wang2025sociobench} for countries in America, Asia, Europe, and Oceania.
For the three African countries, we instead evaluate on WVS using an 80/20 split, due to the lack of comparable ISSP coverage.
To verify that this coverage-driven protocol difference does not drive the main conclusion, Appendix~\ref{sec:appendix_wvs_holdout} reports WVS 80/20 held-out results for the non-African countries as a supplementary robustness check.
We conduct experiments on 14 countries across five continents.\footnote{Country identifiers follow ISO-3166-1 alpha-3. The specific abbreviations and continents are listed in Appendix~\ref{sec:data_details}.}
Appendix~\ref{sec:data_details} reports topic definitions, 13 WVS and 10 ISSP topics, and per-country sample sizes.
We consider four binary persona attributes, namely \textit{Gender}, \textit{Education}, \textit{Residence}, and \textit{Marital Status}, as these attributes are known to systematically relate to cultural values and attitudes in cross-national survey research~\citep{kenny2017gender,lomazzi2020gender,serrano2020individual}.
Country is treated as part of the context $X$.

\paragraph{Baselines.}
Table~\ref{tab:main_res} compares ACE-Align against external baselines and internal ablations, all instantiated on LLaMA3.1-8B-Instruct~\cite{dubey2024llama}.
Additional results on Qwen3-8B~\citep{yang2025qwen3technicalreport} and Llama-3.1-70B-Instruct are reported in Appendix~\ref{sec:appendix_additional_backbones}.
\textbf{(1) Base model:} LLaMA3.1-8B-Instruct is evaluated using the same persona-conditioned prompting protocol, without any additional alignment training.
\textbf{(2) Prompt-based Method:} \textit{Anthropological Prompting} augments prompting with an explicit anthropological reasoning framework before selecting an option~\cite{alkhamissi2024investigating}.
\textbf{(3) SFT-based Methods:} We include \textit{SimLLM}, \textit{CultureSPA}, and the no-augmentation variant of \textit{CultureLLM} as supervised fine-tuning baselines~\cite{cao2025specializing,xu2025self,li2024culturellm}.
\textbf{(4) RL-based Methods:} We include DPO, PPO, and GRPO to test whether standard preference or policy optimization improves cultural alignment without explicit attribute-effect matching~\cite{rafailov2023direct,schulman2017proximal,shao2024deepseekmath}.
\textbf{(5) Ablation Variants:} Anchor-only is equivalent to the no-augmentation CultureLLM variant and optimizes only $\mathcal{L}_{\text{anchor}}$ (Eq.~\ref{eq:overall_obj} with $\beta=0$). Causal-only optimizes only $\mathcal{L}_{\text{CE}}$ (Eq.~\ref{eq:overall_obj} with $\alpha=0$). Together, they isolate the two objectives in ACE-Align.

\paragraph{Evaluation Metric.}
Following \citet{santurkar2023whose}, we evaluate cultural alignment by comparing the model predicted opinion distribution to the reference human distribution on each multiple choice survey question.
For each question $q\in Q$, let $D_1(q)$ and $D_2(q)$ denote two distributions over the answer choices (e.g., model vs.\ survey).
To respect the ordinal structure of survey options, we use the 1-Wasserstein distance after embedding options into a 1D metric space (details below).
We then define the alignment score $S$ between $D_1$ and $D_2$ over a question set $Q$ as
\begin{equation}
S
=
\frac{1}{|Q|}
\sum_{q\in Q}
\left(
1
-
\frac{\mathrm{WD}\!\big(D_1(q), D_2(q)\big)}{N-1}
\right),
\label{eq:wd_align}
\end{equation}
where $N$ is the number of ordinal answer choices, and $N-1$ is the maximum possible Wasserstein distance, ensuring $S\in[0,1]$. For readability, all reported scores are scaled by 100.

\subsection{Experimental Results}

\begin{figure*}[t]
    \centering
    \begin{subfigure}[b]{0.32\textwidth}
        \centering
        \includegraphics[width=\textwidth]{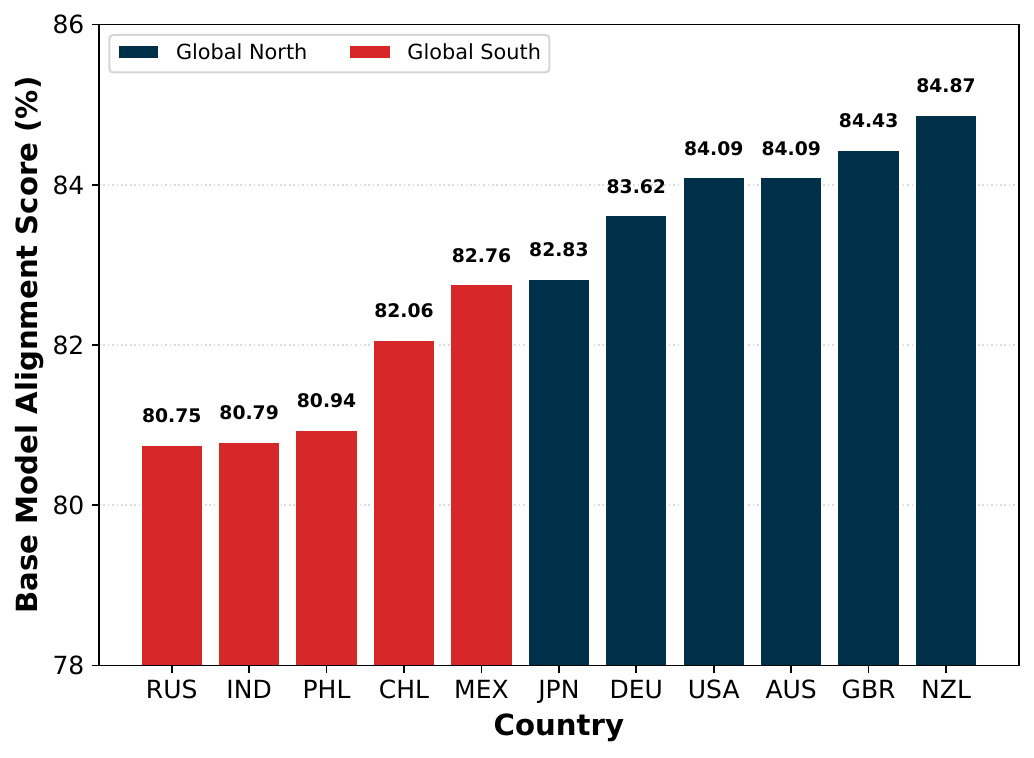}
        \caption{ISSP Base Scores by Country}
        \label{fig:issp_country_base}
    \end{subfigure}
    \hfill
    \begin{subfigure}[b]{0.32\textwidth}
        \centering
        \includegraphics[width=\textwidth]{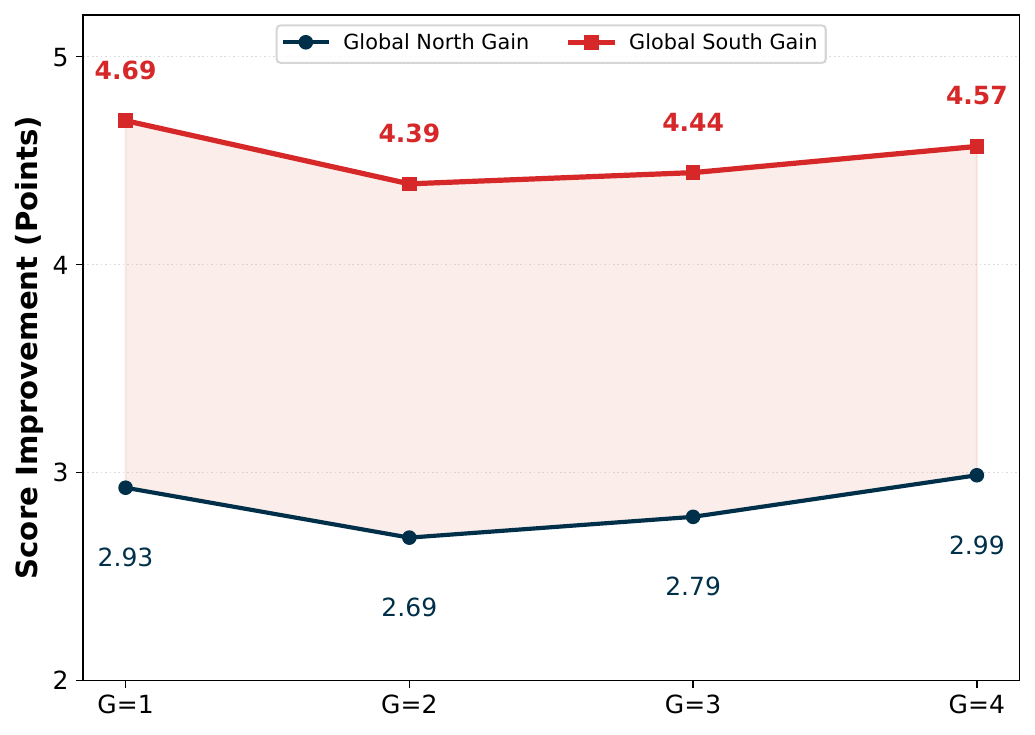}
        \caption{ACE-Align Gains by Group}
        \label{fig:issp_group_gain}
    \end{subfigure}
    \hfill
    \begin{subfigure}[b]{0.32\textwidth}
        \centering
        \includegraphics[width=\textwidth]{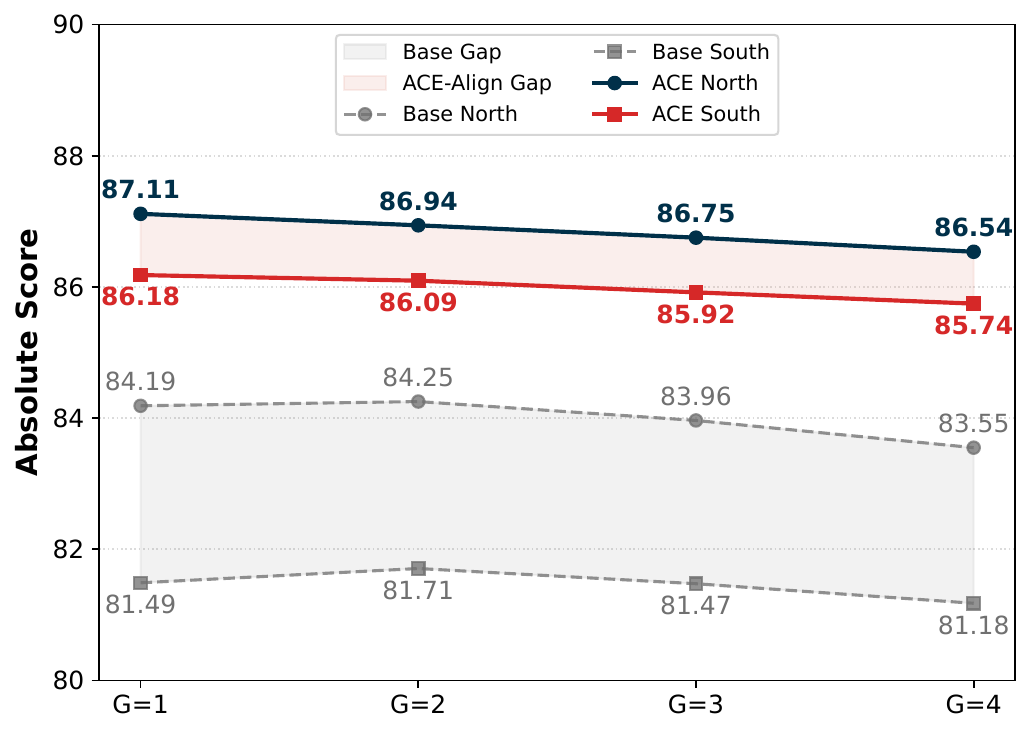}
        \caption{ISSP Group Scores and Gaps}
        \label{fig:issp_group_scores}
    \end{subfigure}

    \caption{\textbf{Within-ISSP geographic equity across persona granularities.}
    (a) Base-model country scores averaged across $G\in\{1,2,3,4\}$, sorted by score and color-coded by geographic group.
    (b) Group-mean gains of ACE-Align over the base model.
    (c) Group-mean absolute scores for the base model and ACE-Align; shaded bands show the directional Global North--South gap, defined as mean(Global North) minus mean(Global South).
    ACE-Align reduces this gap at every granularity.}

    \label{fig:main_results}
\end{figure*}

\paragraph{RQ1: Does ACE-Align consistently improve cultural alignment across varying persona granularities?}
Table~\ref{tab:main_res} shows that ACE-Align achieves the highest average alignment scores across all persona granularities $G\in\{1,2,3,4\}$. Compared to the base model, ACE-Align improves the average score by +4.62 at $G{=}1$, +4.51 at $G{=}2$, +4.66 at $G{=}3$, and +4.82 at $G{=}4$. Country-clustered paired tests in Appendix~\ref{sec:appendix_significance} further compare ACE-Align with all baselines in Table~\ref{tab:main_res}. Although performance generally reduces as $G$ increases, higher granularity creates smaller persona subgroups and more compositional alignment constraints. ACE-Align nevertheless maintains robust gains across all granularities, indicating that its improvements persist from coarse to fine persona granularity. Appendix~\ref{sec:appendix_category_level} further reports category-level alignment patterns across persona granularities.

\begin{figure}[t]
  \centering
  \begin{minipage}[t]{0.48\columnwidth}
    \centering
    \includegraphics[width=\linewidth]{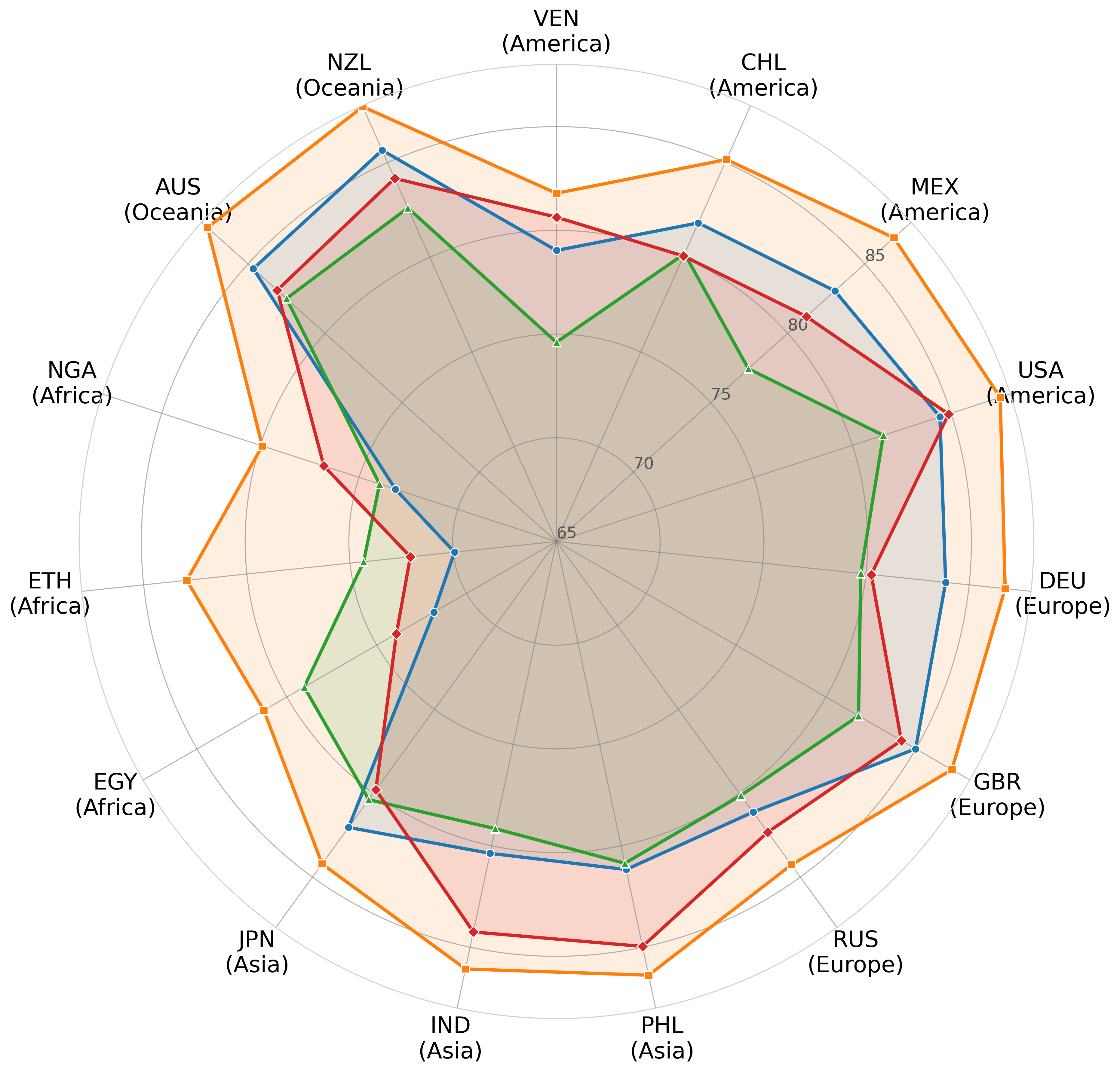}
    \subcaption{Holdout group 1}
    \label{fig:radar_holdout_g1}
  \end{minipage}\hfill
  \begin{minipage}[t]{0.48\columnwidth}
    \centering
    \includegraphics[width=\linewidth]{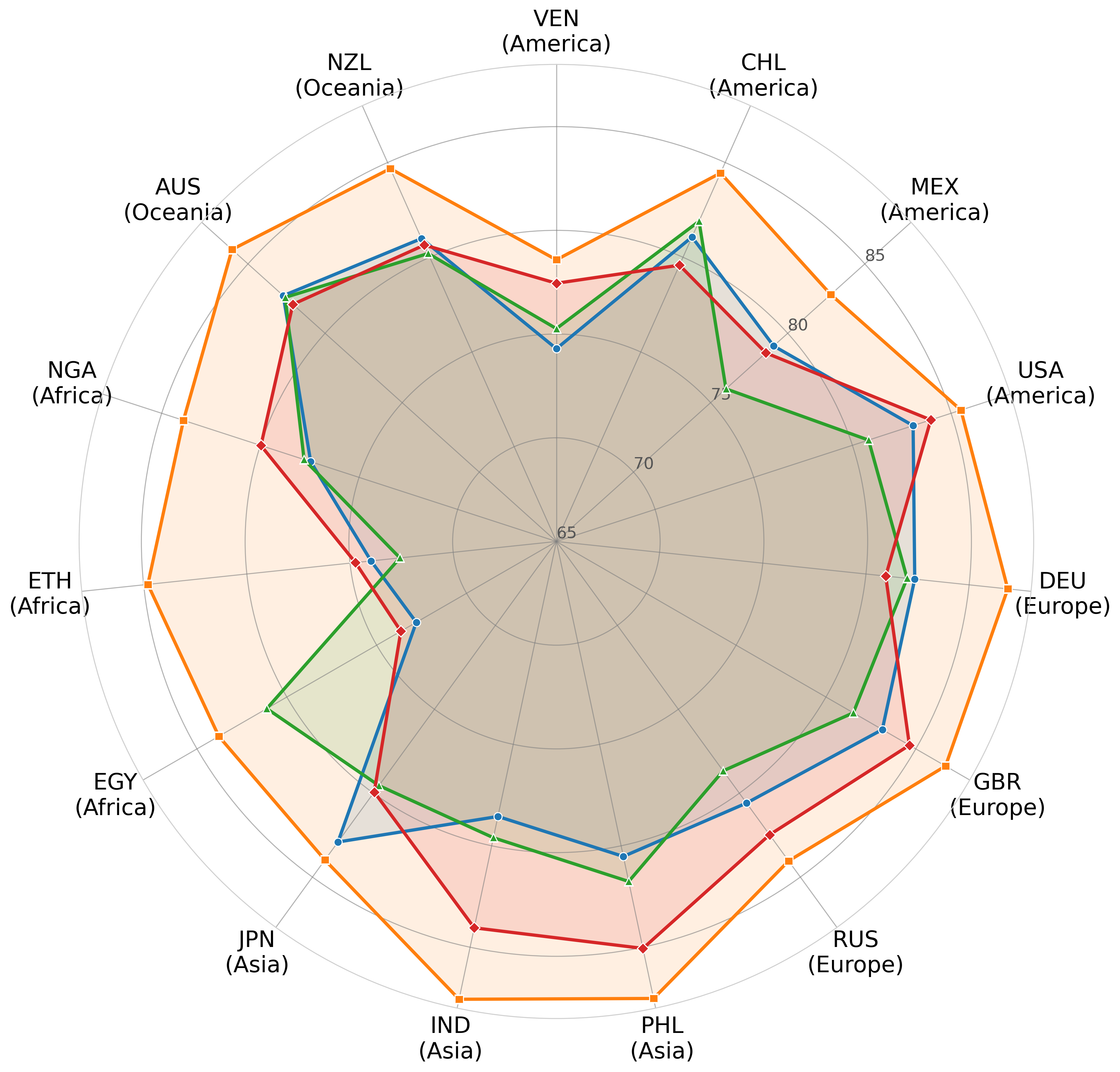}
    \subcaption{Holdout group 2}
    \label{fig:radar_holdout_g2}
  \end{minipage}
    \caption{\textcolor{baseblue}{---Base model} \textcolor{aceorange}{--- ACE-Align}
    \textcolor{culturegreen}{--- Anchor Only}  \textcolor{causalred}{--- Causal Only}
    Alignment on two held-out demographic combinations across cultural themes
    (Female, College Educated, Urban, Married) vs. (Male, Not College Educated, Rural, Not Married).
  }
  \label{fig:radar_holdout}
\end{figure}

\paragraph{RQ2: How robust is ACE-Align under low-resource conditions?}
Low-resource conditions encompass (i) unseen persona compositions and (ii) low-coverage geographic regions.

\textbf{Compositional Generalization.}
As demonstrated in Figure~\ref{fig:radar_holdout}, ACE-Align maintains stable performance on persona profiles whose attribute combinations were withheld during training. This suggests the model captures reusable attribute-effect patterns rather than memorizing training instances.

\textbf{Geographic Equity.}
Following the Brandt-line North--South division~\citep{brandt1980north}, we group USA, DEU, GBR, JPN, AUS, and NZL as the Global North, and CHL, MEX, RUS, IND, PHL, EGY, ETH, and NGA as the Global South. We classify Russia as a Global South emerging power~\citep{jiang2025digital}.

To avoid confounding geographic differences with the evaluation protocol, we compute the Global North--South gap separately within WVS and ISSP. For WVS, all 14 countries are evaluated using the same 80/20 held-out protocol. For ISSP, we compare the six Global North countries with the five covered Global South countries; EGY, ETH, and NGA are excluded because ISSP data are unavailable.

Figure~\ref{fig:main_results} reports the ISSP-only comparison. ACE-Align improves the Global North group by 2.93, 2.69, 2.79, and 2.99 points and the Global South group by 4.69, 4.39, 4.44, and 4.57 points at $G\in\{1,2,3,4\}$, respectively. The corresponding ISSP gap decreases from 2.70, 2.55, 2.49, and 2.37 points for the base model to 0.93, 0.84, 0.83, and 0.79 points for ACE-Align.

Appendix Figure~\ref{fig:wvs_geographic_gap} reports the separate WVS comparison: the gap decreases from 3.38, 3.45, 3.42, and 3.34 points to 1.23, 1.08, 1.02, and 1.09 points. Averaged across granularities, the gap shrinks from 3.40 to 1.11 points on WVS and from 2.53 to 0.85 points on ISSP. Appendix~\ref{sec:appendix_geographic_sensitivity} shows that the conclusion remains unchanged when Russia is reassigned to the Global North.

\paragraph{RQ3: What mechanism drives the effectiveness of ACE-Align?}

To understand \textit{why} ACE-Align works, we analyze the underlying attribute-effect dynamics of cultural attributes. We hypothesize that its effectiveness stems from two key capabilities: (1) correctly identifying that different cultural topics are governed by distinct demographic drivers, and (2) actively correcting specific types of attribute-effect misalignment.

\begin{figure}[t]
    \centering
    \includegraphics[width=\columnwidth]{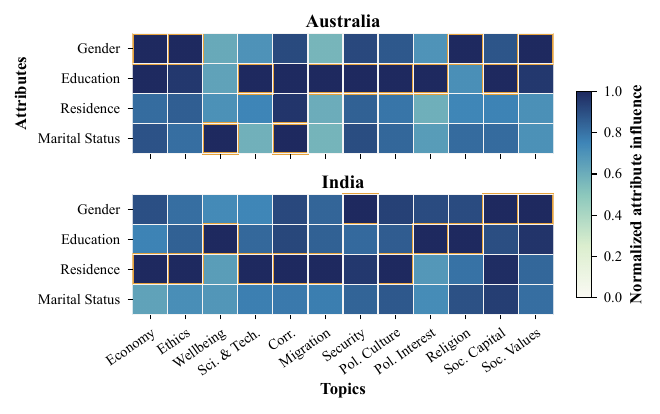}
    \caption{Heterogeneity of Attribute Influence for Australia and India. Different cultural topics and countries are dominated by different demographic attributes. Topic abbreviations are defined in Table~\ref{tab:wvs_topics}.}
    \label{fig:heatmap_attributes}
\end{figure}

\paragraph{Heterogeneity of Attribute Influence.}
Cultural values are not uniformly influenced by all demographics, and the dominant attribute varies substantially across topics and across countries.
As shown in Figure~\ref{fig:heatmap_attributes}, we visualize the survey-grounded attribute-effect magnitude of different attributes across WVS topics.
For \textit{Economic Values}, the dominant attribute is \textit{Gender} in Australia and \textit{Residence} in India.
For \textit{Perceptions of Migration}, it shifts from \textit{Education} in Australia to \textit{Residence} in India.
Appendix~\ref{sec:appendix_attr_heterogeneity} reports the same analysis for the remaining countries.
This supports the design of ACE-Align, which models these variations through the attribute-effect matching loss ($\mathcal{L}_{\text{CE}}$).

\paragraph{Mitigating Stereotyping and Erasure.}
Finally, we analyze how the model aligns these attributes. We categorize the alignment status of each attribute-topic pair into four types based on the relationship between the model's estimated attribute effect ($\Delta_{m}$) and the survey-grounded data effect ($\Delta_{d}$). Table~\ref{tab:alignment_taxonomy} summarizes the resulting taxonomy.

\begin{table}[H]
\centering
\small
\setlength{\tabcolsep}{3pt}
\begin{tabular}{@{}lll@{}}
\toprule
Type & Criterion & Interpretation \\
\midrule
Flipped & $\mathrm{sgn}(\Delta_m) \neq \mathrm{sgn}(\Delta_d)$ & Opposite trend \\
Stereotyping & $|\Delta_m| > |\Delta_d| + \epsilon$ & Overestimated effect \\
Erasure & $|\Delta_m| < |\Delta_d| - \epsilon$ & Underestimated effect \\
Aligned & $|\Delta_m - \Delta_d| \le \epsilon$ & Matched effect \\
\bottomrule
\end{tabular}
\caption{Taxonomy of attribute-effect alignment types, where $\epsilon$ denotes the tolerance.}
\label{tab:alignment_taxonomy}
\end{table}

\begin{figure}[t]
    \centering
    \includegraphics[width=\linewidth]{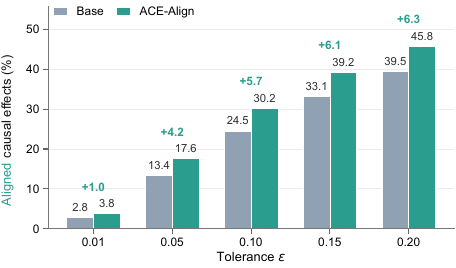}
    \caption{Aligned attribute-effect share across tolerance values $\epsilon$ (USA).}
    \label{fig:overall_comparison}
\end{figure}

Figure~\ref{fig:overall_comparison} shows that ACE-Align increases the share of \textit{Aligned} effects across tolerance values $\epsilon$.
Because this category requires agreement in both direction and magnitude, it directly reflects whether controlled persona edits follow survey-grounded response shifts.
Appendix~\ref{sec:appendix_overall_misalignment} reports category-wise distributions for all countries.
Together, these results show that ACE-Align improves both aggregate alignment and structural attribute-effect consistency.

\section{Conclusion}

We presented ACE-Align, a controlled persona-edit framework for cultural value alignment under varying persona granularities. ACE-Align matches model-side response shifts with survey-grounded attribute effects while anchoring absolute predictions to empirical response modes. Across 14 countries and four persona granularities, ACE-Align improves cultural alignment. In within-survey comparisons, it reduces the average Global North--South alignment gap from 3.40 to 1.11 points on WVS and from 2.53 to 0.85 points on ISSP. These results show that modeling attribute-level response patterns can improve robustness to persona granularity and reduce geographic disparity in cultural alignment.

\section*{Limitations}

There are still some practical limits to our study that we need to address. Because we rely on existing social surveys, we cannot represent every single culture or small community across the globe. The specific data available in these surveys also influenced our study, as we were limited to choosing only four primary traits: gender, education, residence, and marital status. Currently, we treat these traits as binary categories, such as urban versus rural or married versus not married. This binary approach is a simplified version of reality and does not capture the full complexity of how people actually identify themselves. Unobserved demographic factors and interactions among attributes may also affect survey responses, so controlled persona edits should be interpreted as observed-covariate contrasts rather than fully identified real-world causal effects. In the future, we want to look beyond these simple binary pairs and study more multivariate demographic factors. Moving toward these diverse and multi-layered factors will help our models reflect the true diversity of global cultures more fairly.

\section*{Ethics Statement}

This work uses publicly available, de-identified survey datasets and does not collect new human-subject data. ACE-Align models population-level associations between demographic attributes and cultural-value responses. These associations should not be interpreted as deterministic descriptions of individuals or as fully identified real-world causal effects. Improper deployment may reinforce cultural stereotypes, oversimplify demographic groups, or disadvantage populations that are underrepresented in the surveys. Accordingly, the framework is intended for research and diagnostic purposes. Model outputs should be interpreted in context and with appropriate human oversight.

\section*{Acknowledgments}

This work was supported by the Strategic Priority Research Program of the Chinese Academy of Sciences (CAS) under Grant No.~XDB0680302, the Director's Fund Project of the State Key Laboratory of AI Safety, and the Young Elite Scientists Sponsorship Program of the Beijing High Innovation Plan under Grant No.~20250924. We used GPT for polishing the text.

\bibliography{custom}

\appendix

\section{Details of Data Statistics}
\label{sec:data_details}

\subsection{World Values Survey}
\label{sec:wvs}

Tables~\ref{tab:wvs_topics} and~\ref{tab:country_samples} summarize the WVS Wave~7 statistics used in this work, including culture topics, abbreviations, question counts, and country-level respondent counts after preprocessing.

\begin{table}[H]
\centering
\small
\setlength{\tabcolsep}{3pt}
\renewcommand{\arraystretch}{1.05}
\begin{tabular*}{\linewidth}{@{\extracolsep{\fill}}>{\raggedright\arraybackslash}p{0.22\linewidth}>{\raggedright\arraybackslash}p{0.52\linewidth}r@{}}
\toprule
Abbr. & Category & \#Q \\
\midrule
Soc. Values & Social Values, Norms, Stereotypes & 44 \\
Soc. Capital & Social Capital, Trust and Organizational Membership & 36 \\
Pol. Interest & Political Interest and Political Participation & 36 \\
Pol. Culture & Political Culture and Political Regimes & 25 \\
Ethics & Ethical Values & 23 \\
Security & Perceptions of Security & 21 \\
Religion & Religious Values & 12 \\
Wellbeing & Happiness and Wellbeing & 11 \\
Migration & Perceptions of Migration & 10 \\
Corr. & Perceptions of Corruption & 9 \\
Economy & Economic Values & 6 \\
Sci. \& Tech. & Perceptions about Science and Technology & 6 \\
Demo. & Demographic and Socioeconomic Variables & 1 \\
\bottomrule
\end{tabular*}
\caption{WVS topic categories, abbreviations, and question counts.}
\label{tab:wvs_topics}
\end{table}

\begin{table}[H]
\centering
\small
\setlength{\tabcolsep}{8pt}
\renewcommand{\arraystretch}{1.05}
\begin{tabular*}{\linewidth}{@{\extracolsep{\fill}}lllr@{}}
\toprule
Continent & ID & Country & \#Respondents \\
\midrule
\multirow{3}{*}{Americas} & CHL & Chile & 1000 \\
& MEX & Mexico & 1741 \\
& USA & United States & 2596 \\
\addlinespace
\multirow{3}{*}{Europe} & DEU & Germany & 1528 \\
& GBR & Great Britain & 2609 \\
& RUS & Russia & 1810 \\
\addlinespace
\multirow{3}{*}{Asia} & IND & India & 1692 \\
& JPN & Japan & 1353 \\
& PHL & Philippines & 1200 \\
\addlinespace
\multirow{3}{*}{Africa} & EGY & Egypt & 1200 \\
& ETH & Ethiopia & 1230 \\
& NGA & Nigeria & 1237 \\
\addlinespace
\multirow{2}{*}{Oceania} & AUS & Australia & 1813 \\
& NZL & New Zealand & 1057 \\
\bottomrule
\end{tabular*}
\caption{WVS respondents per country after preprocessing. Country identifiers follow ISO-3166-1 alpha-3. Continents are grouped into Americas, Europe, Asia, Africa, and Oceania.}
\label{tab:country_samples}
\end{table}

\subsection{International Social Survey Programme}
\label{sec:issp}

Table~\ref{tab:issp_dimensions} summarizes the ISSP dimensions used for evaluation, and Table~\ref{tab:issp_country_dimension_samples} reports the corresponding country-level sample counts.

\begin{table}[htbp]
\centering
\small
\setlength{\tabcolsep}{3pt}
\renewcommand{\arraystretch}{1.05}
\begin{tabular*}{\linewidth}{@{\extracolsep{\fill}}lr@{}}
\toprule
Dimension & \#Questions \\
\midrule
Work Orientations & 66 \\
Health & 62 \\
Citizenship & 60 \\
Role Of Government & 59 \\
National Identity & 58 \\
Social Networks & 58 \\
Religion & 56 \\
Family & 50 \\
Environment & 49 \\
Social Inequality & 46 \\
\bottomrule
\end{tabular*}
\caption{ISSP dimensions and the number of survey questions per dimension.}
\label{tab:issp_dimensions}
\end{table}

\begin{table*}[t]
\centering
\small
\setlength{\tabcolsep}{6pt}
\renewcommand{\arraystretch}{1.05}
\begin{tabular*}{\textwidth}{@{\extracolsep{\fill}}lrrrrrrrrrrr@{}}
\toprule
Country & Citiz. & Envir. & Family & Health & Nat. Id. & Relig. & Govt. & Ineq. & Nets. & Work & \textbf{Total}  \\
\midrule
Australia      & 1432 & 1147 & 1612 & 1050 & -- & -- & 1267 & 1068 & 1317 & 1211 & 10104 \\
Chile          & 1432 & --   & 1564 & --   & -- & 1402 & 1416 & 1374 & --   & 1433 &  8621 \\
Germany        & 1718 & 1702 & 1766 & 1744 & 1717 & 1724 & 1689 & 1325 & 1701 & 1687 & 16773 \\
Great Britain  & 1580 & --   &  950 & --   &  904 & 1552 & 1563 & 1724 & 1595 & 1793 & 11661 \\
India          & 1209 & 1421 & 1660 & 1683 & 1530 & --   & 1508 & --   & 1510 & 1336 & 11857 \\
Japan          & 1593 & 1491 & 1212 & 1453 & 1234 & 1466 & 1611 & 1473 & 1609 & 1573 & 14715 \\
Mexico         & --   & --   & 1527 & 1001 & 1062 & --   & --   & --   & 1002 & 1141 &  5733 \\
New Zealand    & --   &  993 & --   & 1135 & --   & 1334 & 1350 & 1210 & 1357 &  901 &  8280 \\
Philippines    & 1200 & 1500 & 1200 & 1800 & 1200 & 1200 & 1200 & 4250 & 1200 & 1200 & 15950 \\
Russia         & 1600 & 1583 & 1525 & 1597 & 1516 & 1583 & 1576 & 1597 & 1559 & 1596 & 15732 \\
United States  & 1264 & 1847 & 1302 & 1146 & 1274 & 1175 & 1390 & 1852 & 1173 & 1477 & 13900 \\
\bottomrule
\end{tabular*}
\caption{ISSP sample counts by country and dimension. Dashes indicate that the corresponding module is not available for that country in our data.}
\label{tab:issp_country_dimension_samples}
\end{table*}

\section{Persona Attribute Coding}
\label{sec:attribute_coding}

Table~\ref{tab:attribute_coding} reports the survey-specific mappings used to harmonize the four demographic attributes and the binary coding used to define the direction of each persona-attribute effect.
We treat $A=1$ as the focal attribute value and $A=0$ as the corresponding contrast value, so that estimated effect directions are comparable across countries, cultural topics, and persona granularities.
Missing values, refusals, and ``don't know'' responses are treated as invalid and excluded before binarization.

\begin{table*}[!t]
\centering
\small
\setlength{\tabcolsep}{4pt}
\renewcommand{\arraystretch}{1.10}
\begin{tabular*}{\textwidth}{@{\extracolsep{\fill}}>{\raggedright\arraybackslash}p{0.09\textwidth}>{\raggedright\arraybackslash}p{0.23\textwidth}>{\raggedright\arraybackslash}p{0.34\textwidth}>{\raggedright\arraybackslash}p{0.11\textwidth}>{\raggedright\arraybackslash}p{0.11\textwidth}@{}}
\toprule
\textbf{Attribute} & \textbf{WVS mapping} & \textbf{ISSP mapping} & \boldmath$\mathbf{A=1}$ & \boldmath$\mathbf{A=0}$ \\
\midrule
Gender & Male vs. Female & \texttt{SEX}: 1 as Male and 2 as Female & Female & Male \\
Education & ISCED 0--3 as No College and ISCED 4--8 as College & \texttt{DEGREE}: 1--4 as No College and 5--6 as College. If only \texttt{EDULEVEL} is available, 0--4 and 5--8 are used & Not college educated & College educated \\
Residence & \texttt{H\_URBRURAL}: Urban vs. Rural & \texttt{URBRURAL}: 1--3 as Urban and 4--5 as Rural & Rural & Urban \\
Marital status & Married or living together as married vs. divorced, separated, widowed, or single & \texttt{MARITAL}: 1--3 as Married and 4--6 as Not Married & Not married & Married \\
\bottomrule
\end{tabular*}
\caption{Survey-specific demographic mappings and binary treatment coding used for effect-direction consistency.}
\label{tab:attribute_coding}
\end{table*}

\section{Category-level Results across Persona Granularities}
\label{sec:appendix_category_level}

\begin{figure*}[p]
\centering
\includegraphics[width=\textwidth]{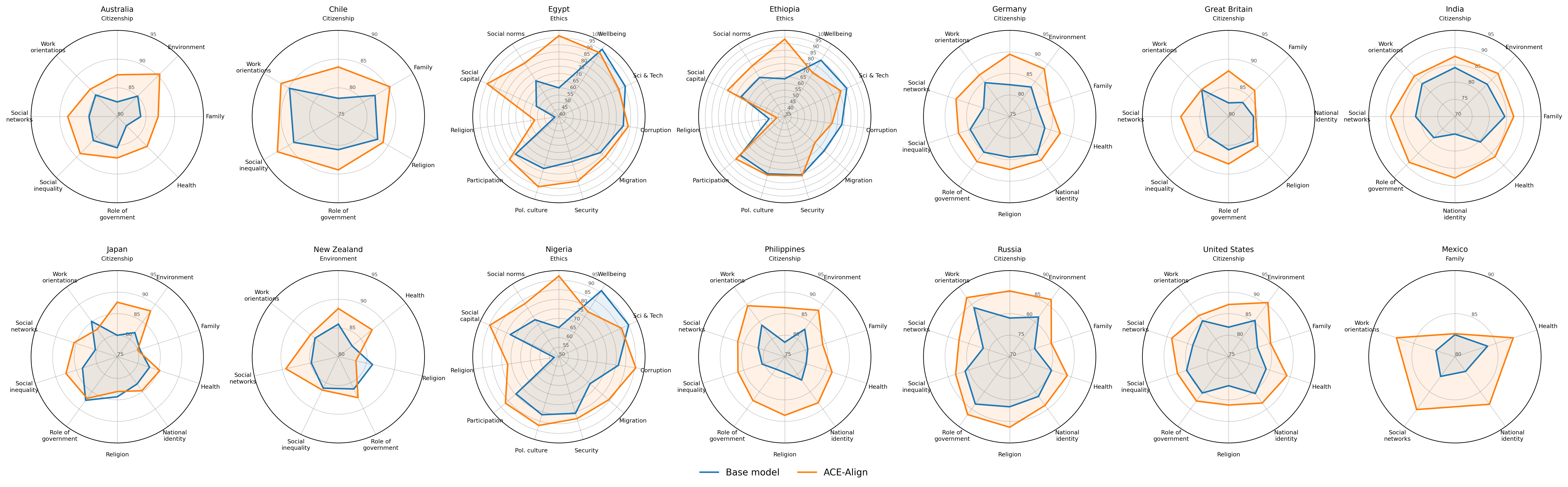}
\caption{Category-level alignment under persona granularity $G=1$.}
\label{fig:category_level_g1}
\end{figure*}

\begin{figure*}[p]
\centering
\includegraphics[width=\textwidth]{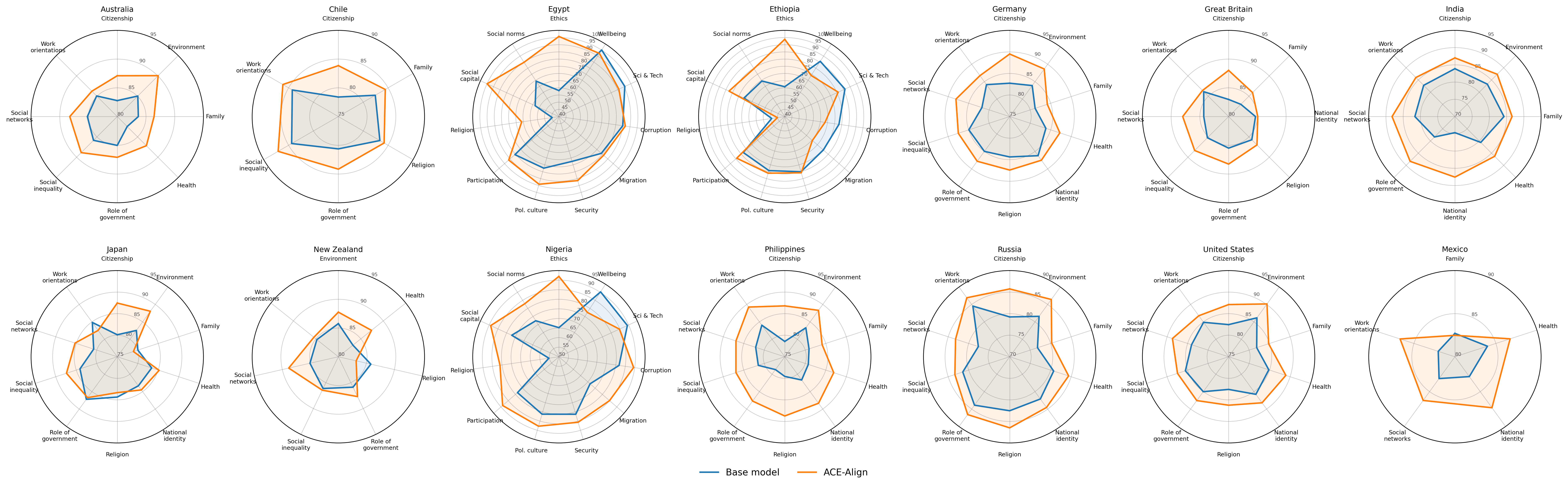}
\caption{Category-level alignment under persona granularity $G=2$.}
\label{fig:category_level_g2}
\end{figure*}

\begin{figure*}[p]
\centering
\includegraphics[width=\textwidth]{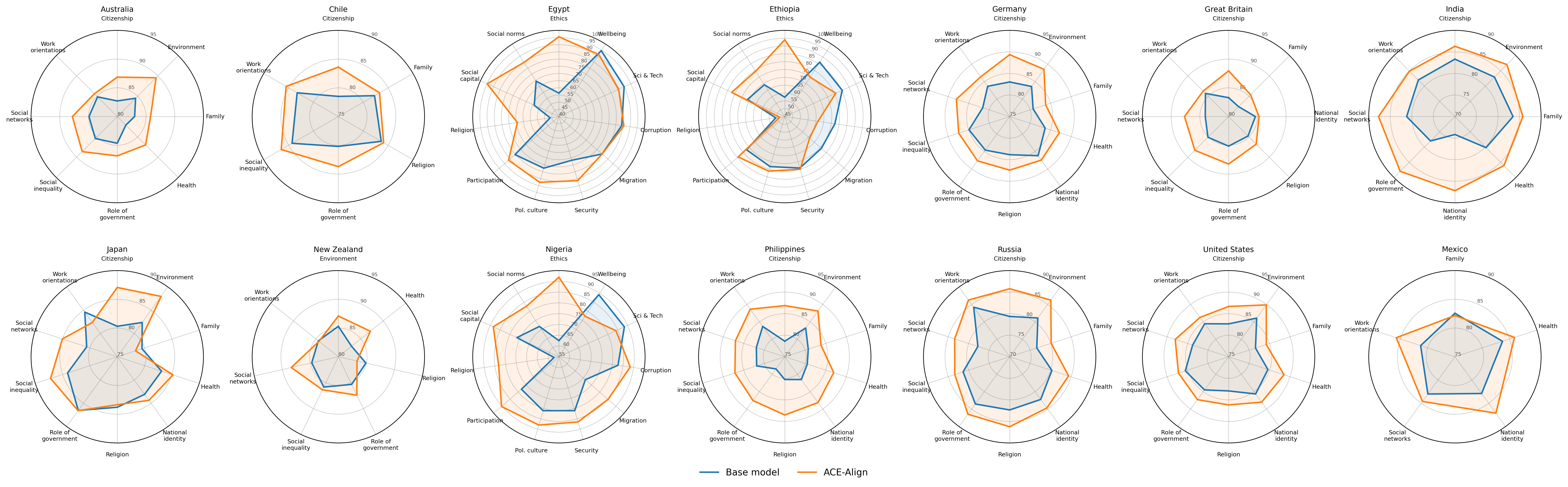}
\caption{Category-level alignment under persona granularity $G=3$.}
\label{fig:category_level_g3}
\end{figure*}

\begin{figure*}[p]
\centering
\includegraphics[width=\textwidth]{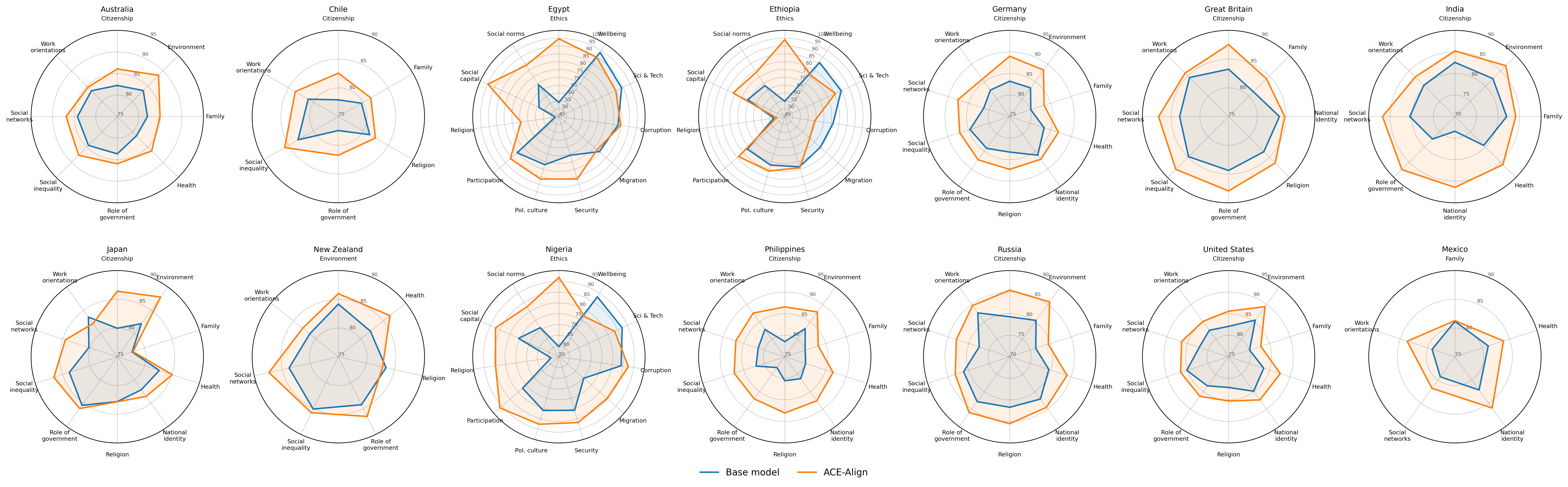}
\caption{Category-level alignment under persona granularity $G=4$.}
\label{fig:category_level_g4}
\end{figure*}

We report category-level cultural alignment across countries under persona granularities $G\in\{1,2,3,4\}$, as illustrated in Figure~\ref{fig:category_level_g1} to Figure~\ref{fig:category_level_g4}.
In each figure, each radar chart corresponds to one country, axes denote cultural categories, and curves compare the base model and ACE-Align.

\section{Heterogeneity of Attribute Influence}
\label{sec:appendix_attr_heterogeneity}

\begin{figure*}[p]
\centering
\includegraphics[width=\textwidth]{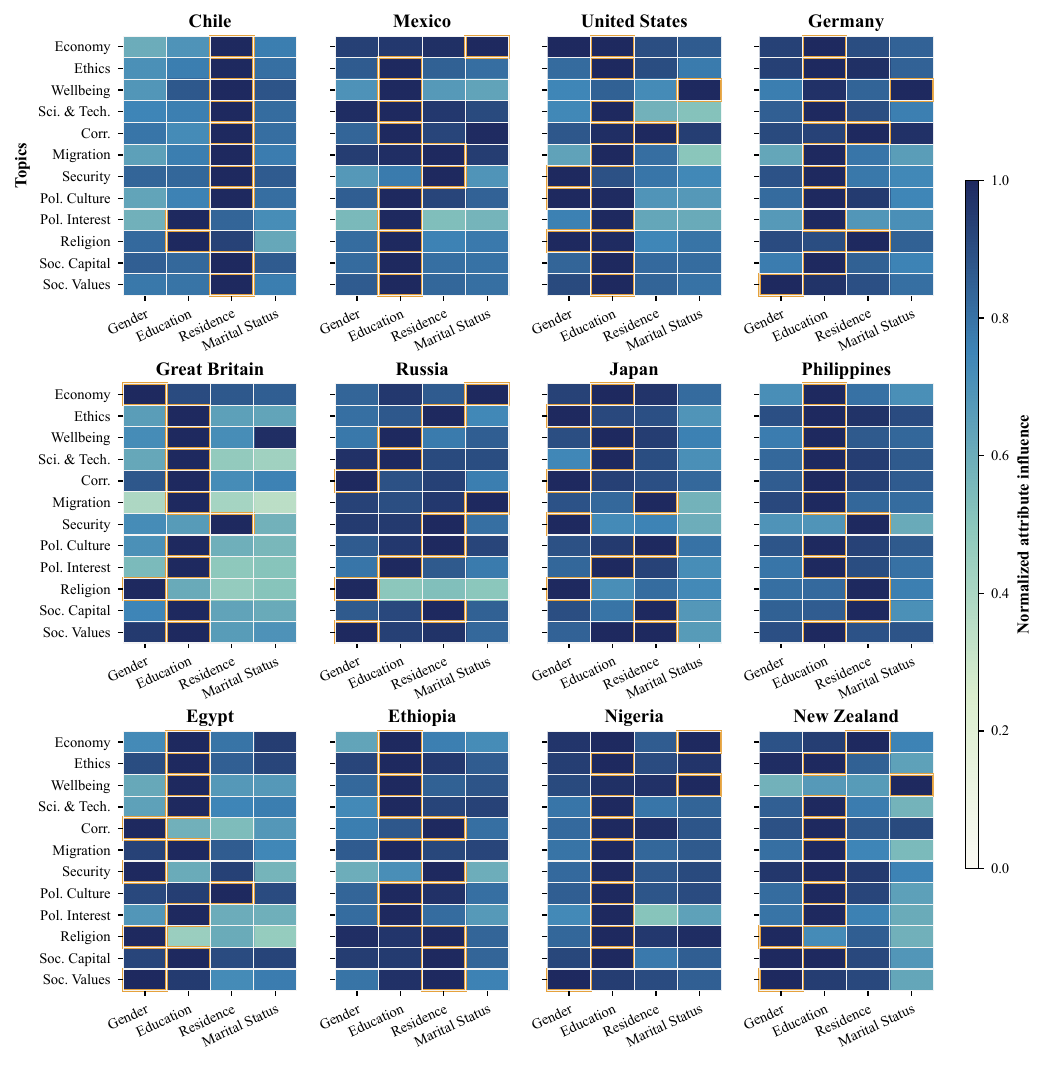}
\caption{Heterogeneity of attribute influence across cultural topics for the remaining countries.
Color intensity indicates the magnitude of the attribute effect. Topic abbreviations are defined in Table~\ref{tab:wvs_topics}.}
\label{fig:attr_influence_remaining}
\end{figure*}

To complement the main analysis, we present attribute influence patterns for the remaining countries not shown in the main text, as illustrated in Figure~\ref{fig:attr_influence_remaining}.
The figure visualizes the relative strength of demographic attributes across cultural topics, highlighting cross-country heterogeneity beyond the illustrative examples discussed earlier.

\section{Overall Comparison of Model Performance and Alignment}
\label{sec:appendix_overall_misalignment}

\begin{figure*}[p]
\centering
\begin{subfigure}[t]{0.245\textwidth}
  \centering
  \includegraphics[width=\linewidth]{Figure/misalignment/panels/United_States/aligned.pdf}
\end{subfigure}\hfill
\begin{subfigure}[t]{0.245\textwidth}
  \centering
  \includegraphics[width=\linewidth]{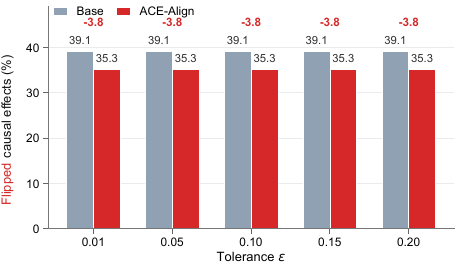}
\end{subfigure}\hfill
\begin{subfigure}[t]{0.245\textwidth}
  \centering
  \includegraphics[width=\linewidth]{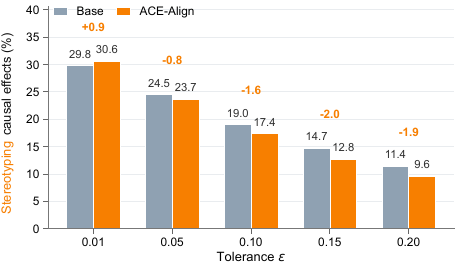}
\end{subfigure}\hfill
\begin{subfigure}[t]{0.245\textwidth}
  \centering
  \includegraphics[width=\linewidth]{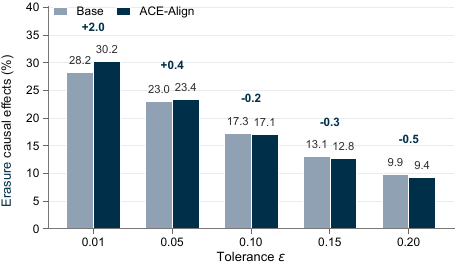}
\end{subfigure}
\caption{Category-wise attribute-effect shares across tolerance values $\epsilon$ (USA).}
\label{fig:overall_misalignment_us}
\end{figure*}

\begin{figure*}[p]
\centering
\begin{subfigure}[t]{0.245\textwidth}
  \centering
  \includegraphics[width=\linewidth]{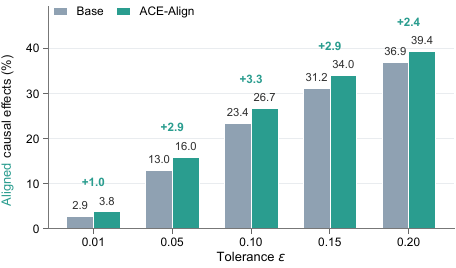}
\end{subfigure}\hfill
\begin{subfigure}[t]{0.245\textwidth}
  \centering
  \includegraphics[width=\linewidth]{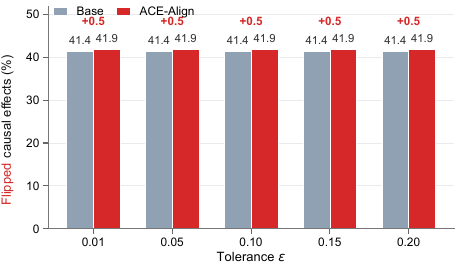}
\end{subfigure}\hfill
\begin{subfigure}[t]{0.245\textwidth}
  \centering
  \includegraphics[width=\linewidth]{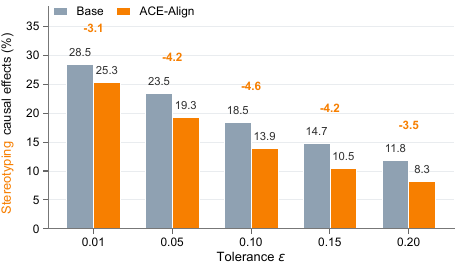}
\end{subfigure}\hfill
\begin{subfigure}[t]{0.245\textwidth}
  \centering
  \includegraphics[width=\linewidth]{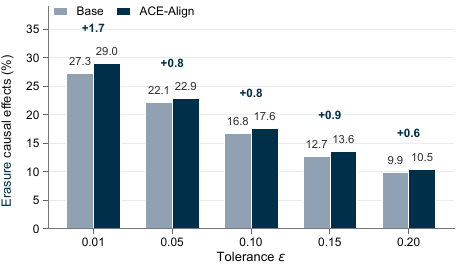}
\end{subfigure}
\caption{Category-wise attribute-effect shares across tolerance values $\epsilon$ (CHL).}
\label{fig:overall_misalignment_chl}
\end{figure*}

\begin{figure*}[p]
\centering
\begin{subfigure}[t]{0.245\textwidth}
  \centering
  \includegraphics[width=\linewidth]{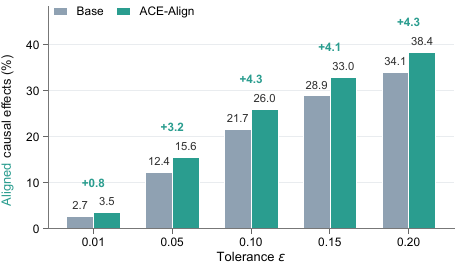}
\end{subfigure}\hfill
\begin{subfigure}[t]{0.245\textwidth}
  \centering
  \includegraphics[width=\linewidth]{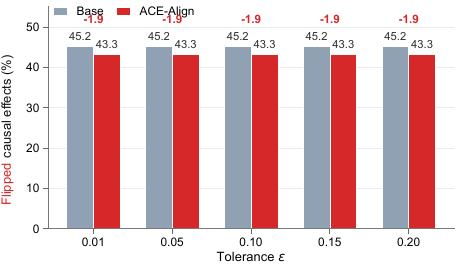}
\end{subfigure}\hfill
\begin{subfigure}[t]{0.245\textwidth}
  \centering
  \includegraphics[width=\linewidth]{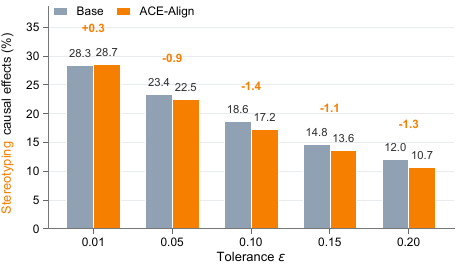}
\end{subfigure}\hfill
\begin{subfigure}[t]{0.245\textwidth}
  \centering
  \includegraphics[width=\linewidth]{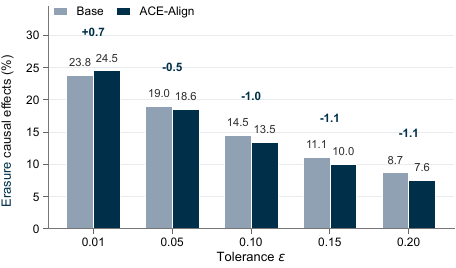}
\end{subfigure}
\caption{Category-wise attribute-effect shares across tolerance values $\epsilon$ (MEX).}
\label{fig:overall_misalignment_mex}
\end{figure*}

\begin{figure*}[p]
\centering
\begin{subfigure}[t]{0.245\textwidth}
  \centering
  \includegraphics[width=\linewidth]{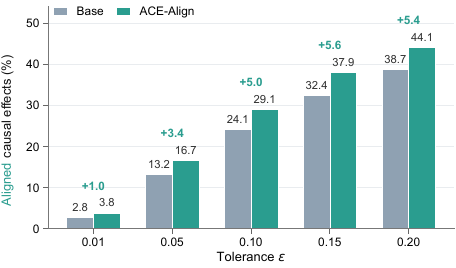}
\end{subfigure}\hfill
\begin{subfigure}[t]{0.245\textwidth}
  \centering
  \includegraphics[width=\linewidth]{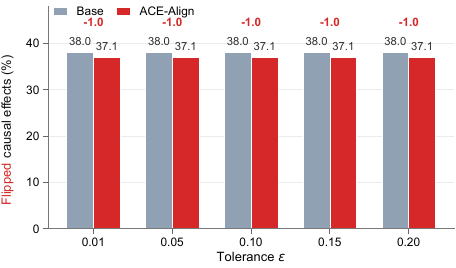}
\end{subfigure}\hfill
\begin{subfigure}[t]{0.245\textwidth}
  \centering
  \includegraphics[width=\linewidth]{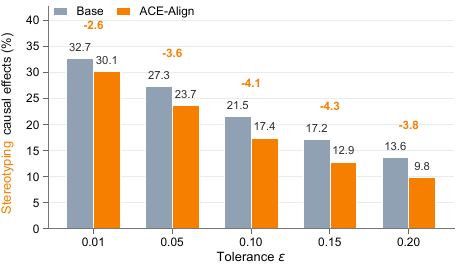}
\end{subfigure}\hfill
\begin{subfigure}[t]{0.245\textwidth}
  \centering
  \includegraphics[width=\linewidth]{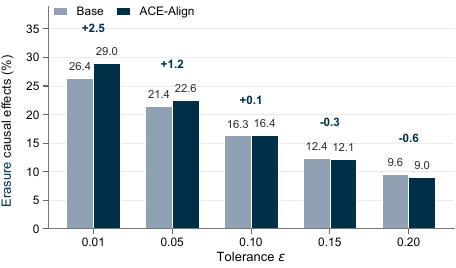}
\end{subfigure}
\caption{Category-wise attribute-effect shares across tolerance values $\epsilon$ (AUS).}
\label{fig:overall_misalignment_aus}
\end{figure*}

\begin{figure*}[p]
\centering
\begin{subfigure}[t]{0.245\textwidth}
  \centering
  \includegraphics[width=\linewidth]{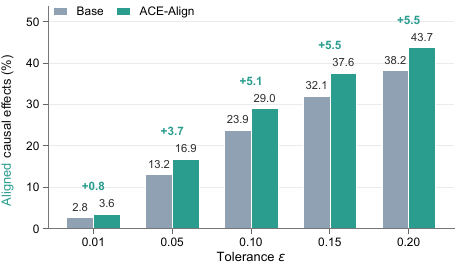}
\end{subfigure}\hfill
\begin{subfigure}[t]{0.245\textwidth}
  \centering
  \includegraphics[width=\linewidth]{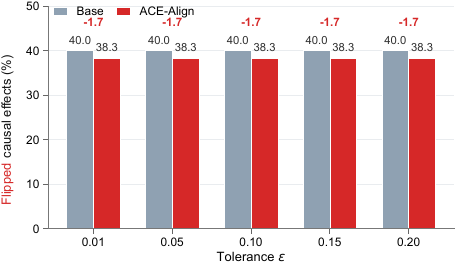}
\end{subfigure}\hfill
\begin{subfigure}[t]{0.245\textwidth}
  \centering
  \includegraphics[width=\linewidth]{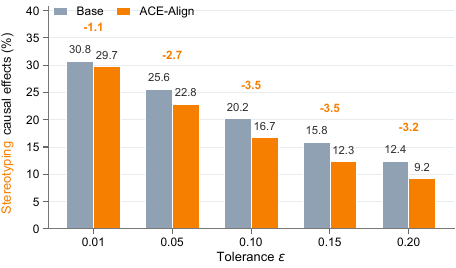}
\end{subfigure}\hfill
\begin{subfigure}[t]{0.245\textwidth}
  \centering
  \includegraphics[width=\linewidth]{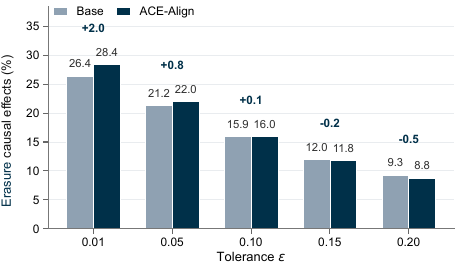}
\end{subfigure}
\caption{Category-wise attribute-effect shares across tolerance values $\epsilon$ (NZL).}
\label{fig:overall_misalignment_nzl}
\end{figure*}

\begin{figure*}[p]
\centering
\begin{subfigure}[t]{0.245\textwidth}
  \centering
  \includegraphics[width=\linewidth]{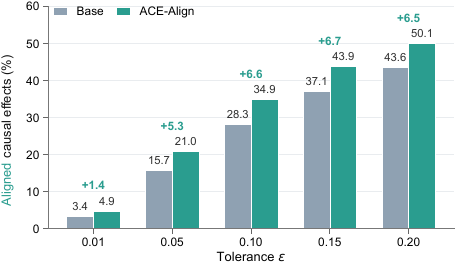}
\end{subfigure}\hfill
\begin{subfigure}[t]{0.245\textwidth}
  \centering
  \includegraphics[width=\linewidth]{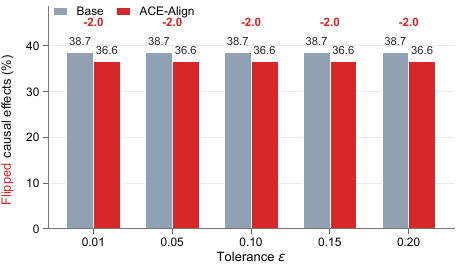}
\end{subfigure}\hfill
\begin{subfigure}[t]{0.245\textwidth}
  \centering
  \includegraphics[width=\linewidth]{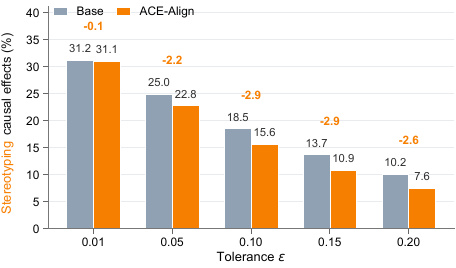}
\end{subfigure}\hfill
\begin{subfigure}[t]{0.245\textwidth}
  \centering
  \includegraphics[width=\linewidth]{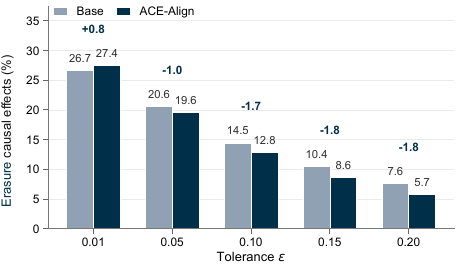}
\end{subfigure}
\caption{Category-wise attribute-effect shares across tolerance values $\epsilon$ (DEU).}
\label{fig:overall_misalignment_deu}
\end{figure*}

\begin{figure*}[p]
\centering
\begin{subfigure}[t]{0.245\textwidth}
  \centering
  \includegraphics[width=\linewidth]{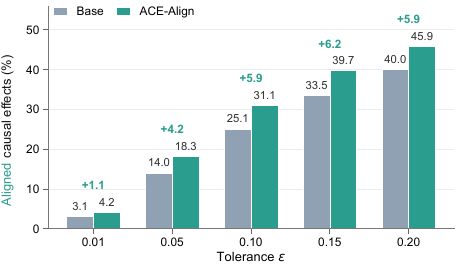}
\end{subfigure}\hfill
\begin{subfigure}[t]{0.245\textwidth}
  \centering
  \includegraphics[width=\linewidth]{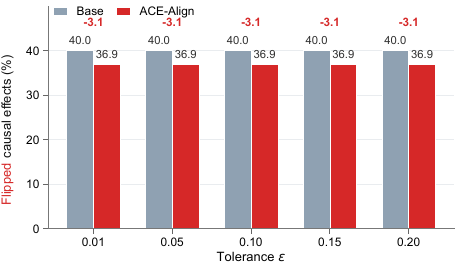}
\end{subfigure}\hfill
\begin{subfigure}[t]{0.245\textwidth}
  \centering
  \includegraphics[width=\linewidth]{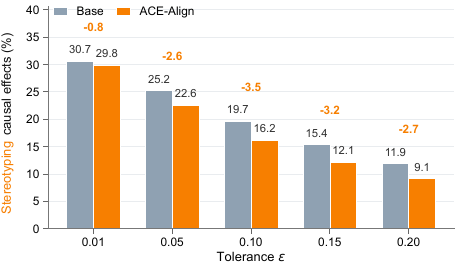}
\end{subfigure}\hfill
\begin{subfigure}[t]{0.245\textwidth}
  \centering
  \includegraphics[width=\linewidth]{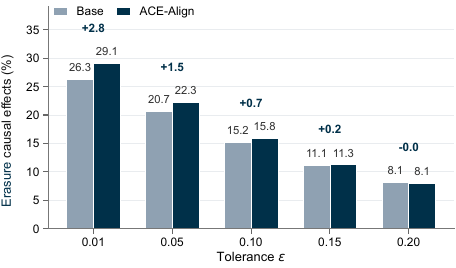}
\end{subfigure}
\caption{Category-wise attribute-effect shares across tolerance values $\epsilon$ (GBR).}
\label{fig:overall_misalignment_gbr}
\end{figure*}

\begin{figure*}[p]
\centering
\begin{subfigure}[t]{0.245\textwidth}
  \centering
  \includegraphics[width=\linewidth]{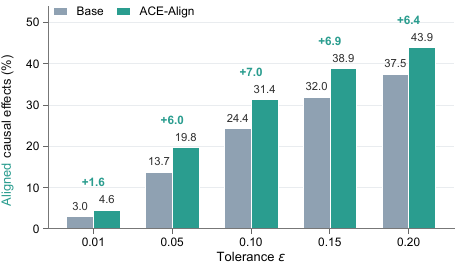}
\end{subfigure}\hfill
\begin{subfigure}[t]{0.245\textwidth}
  \centering
  \includegraphics[width=\linewidth]{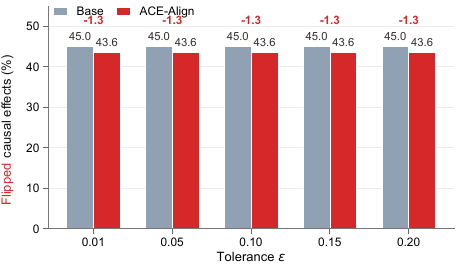}
\end{subfigure}\hfill
\begin{subfigure}[t]{0.245\textwidth}
  \centering
  \includegraphics[width=\linewidth]{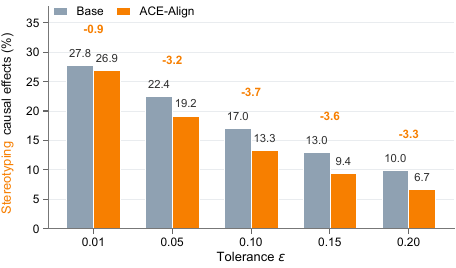}
\end{subfigure}\hfill
\begin{subfigure}[t]{0.245\textwidth}
  \centering
  \includegraphics[width=\linewidth]{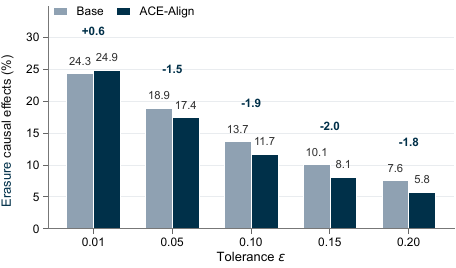}
\end{subfigure}
\caption{Category-wise attribute-effect shares across tolerance values $\epsilon$ (RUS).}
\label{fig:overall_misalignment_rus}
\end{figure*}

\begin{figure*}[p]
\centering
\begin{subfigure}[t]{0.245\textwidth}
  \centering
  \includegraphics[width=\linewidth]{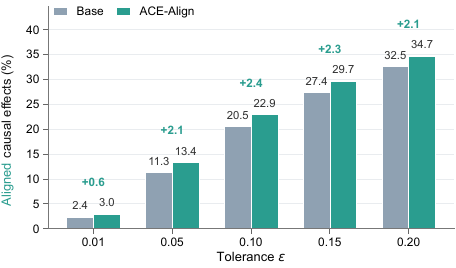}
\end{subfigure}\hfill
\begin{subfigure}[t]{0.245\textwidth}
  \centering
  \includegraphics[width=\linewidth]{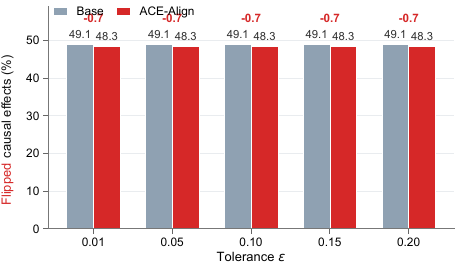}
\end{subfigure}\hfill
\begin{subfigure}[t]{0.245\textwidth}
  \centering
  \includegraphics[width=\linewidth]{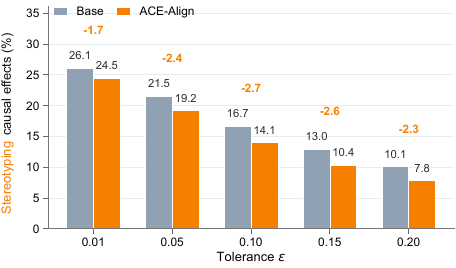}
\end{subfigure}\hfill
\begin{subfigure}[t]{0.245\textwidth}
  \centering
  \includegraphics[width=\linewidth]{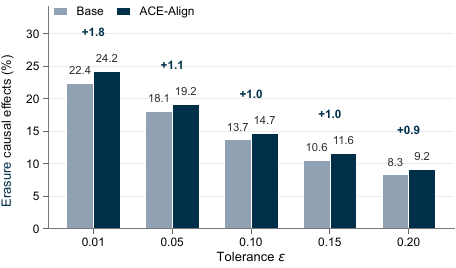}
\end{subfigure}
\caption{Category-wise attribute-effect shares across tolerance values $\epsilon$ (IND).}
\label{fig:overall_misalignment_ind}
\end{figure*}

\begin{figure*}[p]
\centering
\begin{subfigure}[t]{0.245\textwidth}
  \centering
  \includegraphics[width=\linewidth]{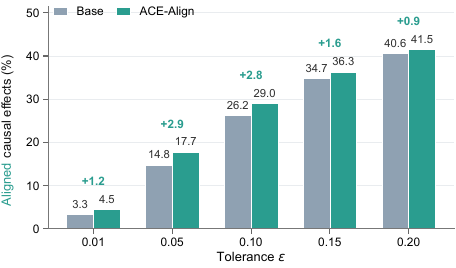}
\end{subfigure}\hfill
\begin{subfigure}[t]{0.245\textwidth}
  \centering
  \includegraphics[width=\linewidth]{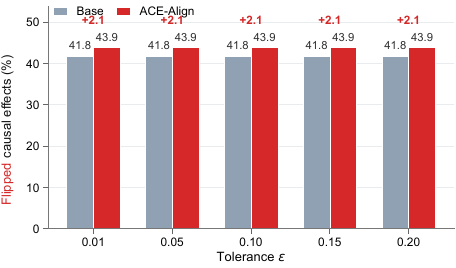}
\end{subfigure}\hfill
\begin{subfigure}[t]{0.245\textwidth}
  \centering
  \includegraphics[width=\linewidth]{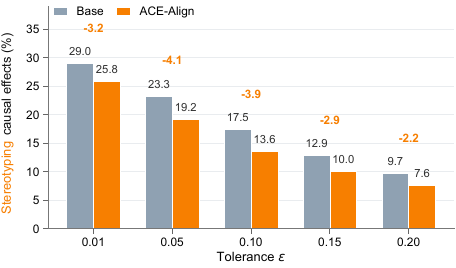}
\end{subfigure}\hfill
\begin{subfigure}[t]{0.245\textwidth}
  \centering
  \includegraphics[width=\linewidth]{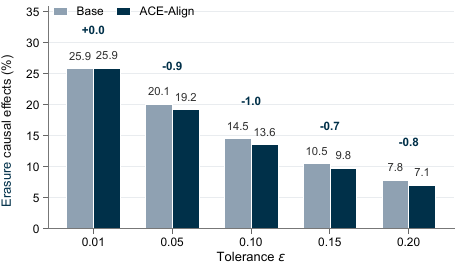}
\end{subfigure}
\caption{Category-wise attribute-effect shares across tolerance values $\epsilon$ (JPN).}
\label{fig:overall_misalignment_jpn}
\end{figure*}

\begin{figure*}[p]
\centering
\begin{subfigure}[t]{0.245\textwidth}
  \centering
  \includegraphics[width=\linewidth]{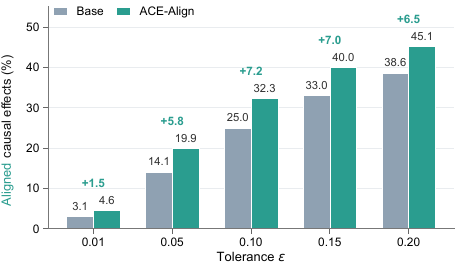}
\end{subfigure}\hfill
\begin{subfigure}[t]{0.245\textwidth}
  \centering
  \includegraphics[width=\linewidth]{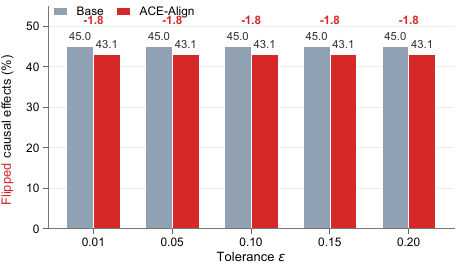}
\end{subfigure}\hfill
\begin{subfigure}[t]{0.245\textwidth}
  \centering
  \includegraphics[width=\linewidth]{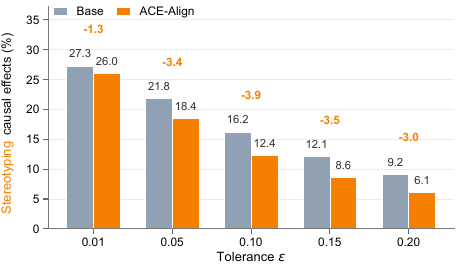}
\end{subfigure}\hfill
\begin{subfigure}[t]{0.245\textwidth}
  \centering
  \includegraphics[width=\linewidth]{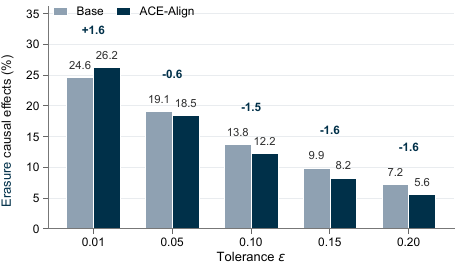}
\end{subfigure}
\caption{Category-wise attribute-effect shares across tolerance values $\epsilon$ (PHL).}
\label{fig:overall_misalignment_phl}
\end{figure*}

\begin{figure*}[p]
\centering
\begin{subfigure}[t]{0.245\textwidth}
  \centering
  \includegraphics[width=\linewidth]{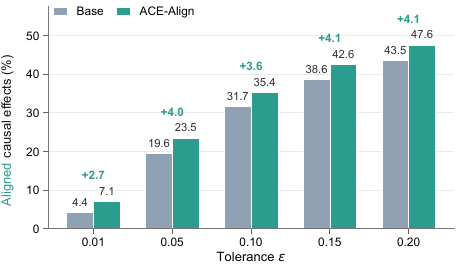}
\end{subfigure}\hfill
\begin{subfigure}[t]{0.245\textwidth}
  \centering
  \includegraphics[width=\linewidth]{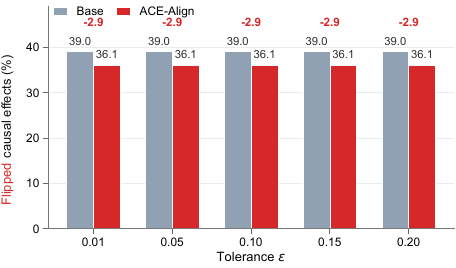}
\end{subfigure}\hfill
\begin{subfigure}[t]{0.245\textwidth}
  \centering
  \includegraphics[width=\linewidth]{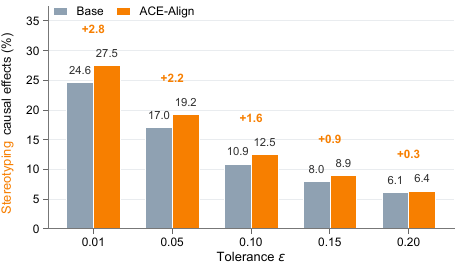}
\end{subfigure}\hfill
\begin{subfigure}[t]{0.245\textwidth}
  \centering
  \includegraphics[width=\linewidth]{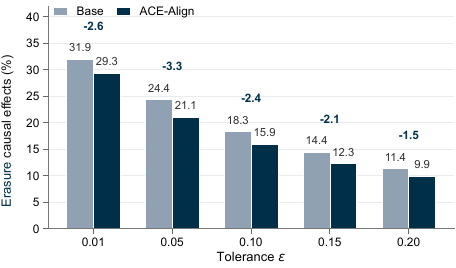}
\end{subfigure}
\caption{Category-wise attribute-effect shares across tolerance values $\epsilon$ (EGY).}
\label{fig:overall_misalignment_egy}
\end{figure*}

\begin{figure*}[p]
\centering
\begin{subfigure}[t]{0.245\textwidth}
  \centering
  \includegraphics[width=\linewidth]{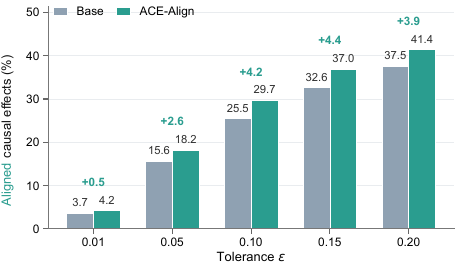}
\end{subfigure}\hfill
\begin{subfigure}[t]{0.245\textwidth}
  \centering
  \includegraphics[width=\linewidth]{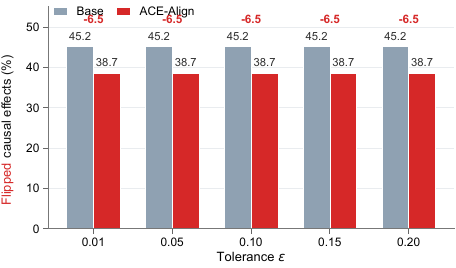}
\end{subfigure}\hfill
\begin{subfigure}[t]{0.245\textwidth}
  \centering
  \includegraphics[width=\linewidth]{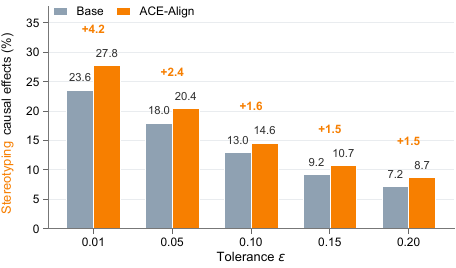}
\end{subfigure}\hfill
\begin{subfigure}[t]{0.245\textwidth}
  \centering
  \includegraphics[width=\linewidth]{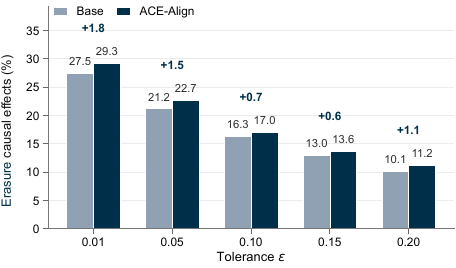}
\end{subfigure}
\caption{Category-wise attribute-effect shares across tolerance values $\epsilon$ (ETH).}
\label{fig:overall_misalignment_eth}
\end{figure*}

\begin{figure*}[p]
\centering
\begin{subfigure}[t]{0.245\textwidth}
  \centering
  \includegraphics[width=\linewidth]{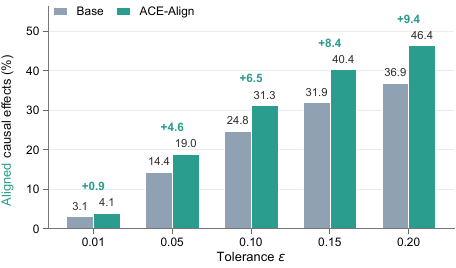}
\end{subfigure}\hfill
\begin{subfigure}[t]{0.245\textwidth}
  \centering
  \includegraphics[width=\linewidth]{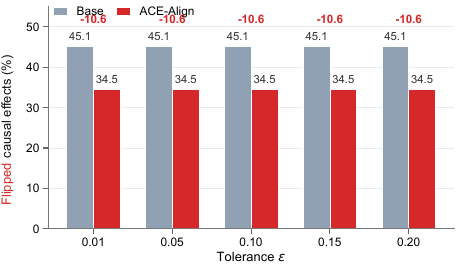}
\end{subfigure}\hfill
\begin{subfigure}[t]{0.245\textwidth}
  \centering
  \includegraphics[width=\linewidth]{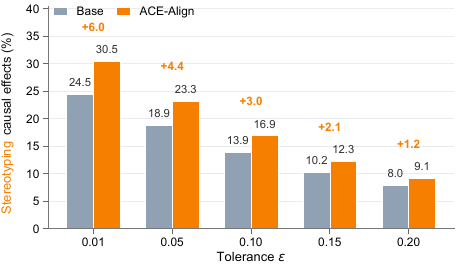}
\end{subfigure}\hfill
\begin{subfigure}[t]{0.245\textwidth}
  \centering
  \includegraphics[width=\linewidth]{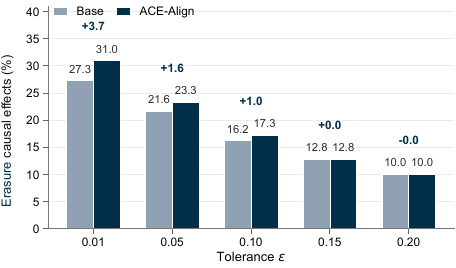}
\end{subfigure}
\caption{Category-wise attribute-effect shares across tolerance values $\epsilon$ (NGA).}
\label{fig:overall_misalignment_nga}
\end{figure*}

Figures~\ref{fig:overall_misalignment_us}--\ref{fig:overall_misalignment_nga} report category-wise attribute-effect alignment across countries.
Each panel shows the share of attribute-effect pairs assigned to one alignment category across tolerance values $\epsilon$.
Colors indicate \textcolor{barbasegray}{Base model}, \textcolor{baraligned}{Aligned}, \textcolor{barflipped}{Flipped}, \textcolor{barstereotyping}{Stereotyping}, and \textcolor{barerasure}{Erasure}.

\section{Implementation Details}
\label{sec:appendix_implementation}

We perform parameter-efficient fine-tuning with Low-Rank Adaptation (LoRA)~\cite{hu2022lora}. We use rank $r=8$, scaling factor $\alpha=16$, and LoRA dropout of $0.05$. Optimization is done with AdamW using a learning rate of $2\times10^{-5}$. We train on two NVIDIA A800 GPUs and use bfloat16 mixed precision.

We train for two epochs with a sequential objective schedule. In the first epoch, we optimize only the anchoring objective in Eq.~\ref{eq:overall_obj} by setting $\alpha=1.0$ and $\beta=0.0$. In the second epoch, starting from the first-epoch checkpoint, we optimize only the effect-alignment objective by setting $\alpha=0.0$ and $\beta=1.0$. All other trained models follow the same optimization and hardware configuration unless stated otherwise.

\section{Additional Backbone Results}
\label{sec:appendix_additional_backbones}

We additionally evaluate ACE-Align on Qwen3-8B, a similarly sized model from a different family, and Llama-3.1-70B-Instruct, a substantially larger model from the Llama family. Tables~\ref{tab:qwen3_country_results} and~\ref{tab:llama70b_country_results} report the complete country-level results. ACE-Align improves the macro-average over the corresponding base model at every persona granularity on both additional backbones.

\begin{table*}[t]
\centering
\small
\setlength{\tabcolsep}{3pt}
\renewcommand{\arraystretch}{1.08}
\resizebox{\textwidth}{!}{
\begin{tabular}{ll|ccc|ccc|ccc|ccc|cc|c}
\toprule
{} & {} & \multicolumn{3}{c|}{\textbf{America}} & \multicolumn{3}{c|}{\textbf{Europe}} & \multicolumn{3}{c|}{\textbf{Asia}} & \multicolumn{3}{c|}{\textbf{Africa}} & \multicolumn{2}{c|}{\textbf{Oceania}} & \textbf{Avg} \\
\cline{3-16}
\textbf{G} & \textbf{Model} & \textbf{CHL} & \textbf{MEX} & \textbf{USA} & \textbf{DEU} & \textbf{GBR} & \textbf{RUS} & \textbf{IND} & \textbf{JPN} & \textbf{PHL} & \textbf{EGY} & \textbf{ETH} & \textbf{NGA} & \textbf{AUS} & \textbf{NZL} & {} \\
\midrule
\multirow{2}{*}{$G=1$}
& Base model & 80.78 & 80.24 & 81.13 & 81.17 & 81.24 & 79.01 & 78.03 & 78.76 & 79.30 & 67.63 & 71.88 & 73.78 & 81.42 & 82.19 & 78.33 \\
& \textbf{ACE-Align} & \textbf{82.41} & \textbf{86.29} & \textbf{85.71} & \textbf{85.19} & \textbf{84.03} & \textbf{82.57} & \textbf{87.21} & \textbf{80.19} & \textbf{84.66} & \textbf{80.20} & \textbf{77.96} & \textbf{84.72} & \textbf{84.69} & \textbf{83.74} & \textbf{83.54} \scriptsize{(+5.21)} \\
\midrule
\multirow{2}{*}{$G=2$}
& Base model & 80.75 & 79.98 & 81.19 & 81.43 & 81.69 & 79.31 & 78.34 & 79.58 & 79.72 & 69.24 & 72.32 & 74.28 & 81.57 & 82.40 & 78.70 \\
& \textbf{ACE-Align} & \textbf{82.39} & \textbf{85.61} & \textbf{85.62} & \textbf{85.22} & \textbf{84.07} & \textbf{82.60} & \textbf{86.92} & \textbf{80.17} & \textbf{84.61} & \textbf{80.49} & \textbf{77.62} & \textbf{84.97} & \textbf{84.57} & \textbf{83.86} & \textbf{83.48} \scriptsize{(+4.78)} \\
\midrule
\multirow{2}{*}{$G=3$}
& Base model & 80.81 & 80.02 & 81.17 & 81.67 & 81.92 & 79.47 & 78.36 & 80.03 & 79.87 & 70.55 & 72.47 & 74.33 & 81.64 & 82.39 & 78.91 \\
& \textbf{ACE-Align} & \textbf{82.45} & \textbf{85.22} & \textbf{85.44} & \textbf{85.24} & \textbf{84.11} & \textbf{82.49} & \textbf{86.24} & \textbf{80.09} & \textbf{84.40} & \textbf{80.79} & \textbf{77.30} & \textbf{84.42} & \textbf{84.40} & \textbf{83.84} & \textbf{83.32} \scriptsize{(+4.41)} \\
\midrule
\multirow{2}{*}{$G=4$}
& Base model & 80.85 & 80.08 & 81.04 & 81.68 & 81.99 & 79.51 & 78.34 & 80.27 & 79.88 & 72.09 & 72.07 & 74.29 & 81.56 & 82.32 & 79.00 \\
& \textbf{ACE-Align} & \textbf{82.55} & \textbf{85.14} & \textbf{85.30} & \textbf{85.20} & \textbf{84.02} & \textbf{82.44} & \textbf{85.25} & 79.95 & \textbf{84.07} & \textbf{80.78} & \textbf{76.76} & \textbf{83.17} & \textbf{84.19} & \textbf{83.90} & \textbf{83.05} \scriptsize{(+4.05)} \\
\bottomrule
\end{tabular}
}
\caption{Country-level cultural alignment scores for Qwen3-8B. Parenthesized values report the macro-average improvement of ACE-Align over the corresponding base model.}
\label{tab:qwen3_country_results}
\end{table*}

\begin{table*}[t]
\centering
\small
\setlength{\tabcolsep}{3pt}
\renewcommand{\arraystretch}{1.08}
\resizebox{\textwidth}{!}{
\begin{tabular}{ll|ccc|ccc|ccc|ccc|cc|c}
\toprule
{} & {} & \multicolumn{3}{c|}{\textbf{America}} & \multicolumn{3}{c|}{\textbf{Europe}} & \multicolumn{3}{c|}{\textbf{Asia}} & \multicolumn{3}{c|}{\textbf{Africa}} & \multicolumn{2}{c|}{\textbf{Oceania}} & \textbf{Avg} \\
\cline{3-16}
\textbf{G} & \textbf{Model} & \textbf{CHL} & \textbf{MEX} & \textbf{USA} & \textbf{DEU} & \textbf{GBR} & \textbf{RUS} & \textbf{IND} & \textbf{JPN} & \textbf{PHL} & \textbf{EGY} & \textbf{ETH} & \textbf{NGA} & \textbf{AUS} & \textbf{NZL} & {} \\
\midrule
\multirow{2}{*}{$G=1$}
& Base model & 82.00 & 80.76 & 84.24 & 84.05 & 84.07 & 82.00 & 80.75 & 83.32 & 81.55 & 76.53 & 75.43 & 76.82 & 83.14 & 84.18 & 81.35 \\
& \textbf{ACE-Align} & \textbf{87.38} & \textbf{88.04} & \textbf{90.05} & \textbf{89.78} & \textbf{90.23} & \textbf{87.71} & \textbf{87.42} & \textbf{83.55} & \textbf{88.27} & \textbf{82.69} & \textbf{82.00} & \textbf{84.61} & \textbf{89.97} & \textbf{88.72} & \textbf{87.17} \scriptsize{(+5.82)} \\
\midrule
\multirow{2}{*}{$G=2$}
& Base model & 81.99 & 80.30 & 83.46 & 83.92 & 83.73 & 81.74 & 80.37 & 83.23 & 81.56 & 76.97 & 74.89 & 76.85 & 82.93 & 83.81 & 81.12 \\
& \textbf{ACE-Align} & \textbf{87.22} & \textbf{87.39} & \textbf{89.77} & \textbf{89.79} & \textbf{90.07} & \textbf{87.52} & \textbf{87.16} & \textbf{83.43} & \textbf{88.14} & \textbf{82.64} & \textbf{81.91} & \textbf{84.02} & \textbf{89.84} & \textbf{88.60} & \textbf{86.96} \scriptsize{(+5.84)} \\
\midrule
\multirow{2}{*}{$G=3$}
& Base model & 81.84 & 80.07 & 82.94 & 83.71 & 83.49 & 81.41 & 80.05 & 83.00 & 81.59 & 77.16 & 74.36 & 76.62 & 82.66 & 83.34 & 80.87 \\
& \textbf{ACE-Align} & \textbf{87.05} & \textbf{87.08} & \textbf{89.36} & \textbf{89.70} & \textbf{89.80} & \textbf{87.24} & \textbf{86.73} & \textbf{83.31} & \textbf{87.95} & \textbf{82.53} & \textbf{81.59} & \textbf{83.56} & \textbf{89.56} & \textbf{88.31} & \textbf{86.70} \scriptsize{(+5.83)} \\
\midrule
\multirow{2}{*}{$G=4$}
& Base model & 81.74 & 80.18 & 82.26 & 83.32 & 83.16 & 81.03 & 79.55 & 82.69 & 81.42 & 77.49 & 73.74 & 76.39 & 82.24 & 82.80 & 80.57 \\
& \textbf{ACE-Align} & \textbf{86.83} & \textbf{87.25} & \textbf{89.09} & \textbf{89.46} & \textbf{89.34} & \textbf{86.94} & \textbf{86.08} & \textbf{83.19} & \textbf{87.61} & \textbf{82.37} & \textbf{81.12} & \textbf{82.87} & \textbf{89.20} & \textbf{88.11} & \textbf{86.39} \scriptsize{(+5.82)} \\
\bottomrule
\end{tabular}
}
\caption{Country-level cultural alignment scores for Llama-3.1-70B-Instruct. Parenthesized values report the macro-average improvement of ACE-Align over the corresponding base model.}
\label{tab:llama70b_country_results}
\end{table*}

Table~\ref{tab:qwen3_baselines} further compares ACE-Align with the evaluated baselines and ablations on Qwen3-8B. ACE-Align achieves the highest macro-average score at every granularity.

\begin{table}[t]
\centering
\small
\setlength{\tabcolsep}{4.5pt}
\renewcommand{\arraystretch}{1.05}
\begin{tabular}{lrrrrr}
\toprule
Method & $G=1$ & $G=2$ & $G=3$ & $G=4$ & Avg. \\
\midrule
Base & 78.33 & 78.70 & 78.91 & 79.00 & 78.74 \\
AdPrompt & 73.63 & 75.09 & 76.10 & 76.75 & 75.39 \\
CultureSPA & 76.56 & 76.77 & 76.77 & 76.81 & 76.73 \\
DPO & 77.84 & 78.25 & 78.50 & 78.57 & 78.29 \\
PPO & 76.16 & 76.48 & 76.74 & 76.75 & 76.53 \\
GRPO & 77.90 & 78.35 & 78.60 & 78.55 & 78.35 \\
Anchor-only & 79.51 & 79.59 & 79.37 & 79.02 & 79.37 \\
Causal-only & 82.11 & 81.99 & 81.70 & 81.29 & 81.77 \\
\textbf{ACE-Align} & \textbf{83.54} & \textbf{83.48} & \textbf{83.32} & \textbf{83.05} & \textbf{83.35} \\
\bottomrule
\end{tabular}
\caption{Macro-average baseline comparison on Qwen3-8B.}
\label{tab:qwen3_baselines}
\end{table}

Finally, we quantify uncertainty in the improvement over each corresponding base model by resampling the 14 countries with replacement 10,000 times while retaining the paired ACE-Align/Base observations. As shown in Table~\ref{tab:additional_backbone_cis}, every paired 95\% bootstrap confidence interval excludes zero.

\begin{table}[t]
\centering
\small
\setlength{\tabcolsep}{3pt}
\renewcommand{\arraystretch}{1.05}
\begin{tabular*}{\linewidth}{@{\extracolsep{\fill}}lcr@{}}
\toprule
Backbone & $G$ & Improvement [95\% CI] \\
\midrule
\multirow{4}{*}{Llama-3.1-8B} & 1 & +4.62 [3.43, 6.11] \\
& 2 & +4.51 [3.23, 6.09] \\
& 3 & +4.66 [3.36, 6.23] \\
& 4 & +4.82 [3.56, 6.36] \\
\midrule
\multirow{4}{*}{Qwen3-8B} & 1 & +5.21 [3.58, 7.09] \\
& 2 & +4.78 [3.22, 6.56] \\
& 3 & +4.41 [2.95, 6.06] \\
& 4 & +4.05 [2.77, 5.45] \\
\midrule
\multirow{4}{*}{Llama-3.1-70B} & 1 & +5.83 [4.84, 6.58] \\
& 2 & +5.84 [4.85, 6.55] \\
& 3 & +5.82 [4.85, 6.52] \\
& 4 & +5.82 [4.86, 6.51] \\
\bottomrule
\end{tabular*}
\caption{Macro-average improvements of ACE-Align over the corresponding base model, with paired country-bootstrap 95\% confidence intervals.}
\label{tab:additional_backbone_cis}
\end{table}

\subsection{RL Baseline Training Signals}
\label{sec:appendix_rl_baselines}

For all RL-based baselines, we use the same backbone model, persona-conditioned prompt format, training split, and evaluation pipeline as ACE-Align.
The supervision signal is derived from the empirical survey response distribution.
For each survey question $x$, let $\mathcal{Y}(x)$ denote the set of valid answer options, and let $p_{\mathrm{data}}(y \mid x)$ be the empirical response distribution estimated from human survey responses.
Model generations are parsed into a discrete answer option $y \in \mathcal{Y}(x)$ by extracting the first valid integer option from the response.

\paragraph{DPO preference pairs.}
For the DPO baseline, we construct preference pairs directly from the empirical survey distribution.
For each survey question $x$, we enumerate pairs of valid answer options $(y_i,y_j)$.
If $p_{\mathrm{data}}(y_i \mid x) > p_{\mathrm{data}}(y_j \mid x)$, option $y_i$ is treated as the preferred response and $y_j$ as the rejected response.
Pairs with equal empirical probabilities are omitted.
Both preferred and rejected responses are formatted as the corresponding discrete option index, using the same answer-parsing convention as evaluation.

\paragraph{PPO reward.}
For the PPO baseline, we use a pointwise reward equal to the empirical probability of the selected answer:
\begin{equation}
r_{\mathrm{PPO}}(x,y)
=
p_{\mathrm{data}}(y \mid x).
\end{equation}
If the model output cannot be parsed into a valid answer option, the reward is set to zero.
This reward assigns larger scores to answers that are more common in the empirical survey distribution.

\paragraph{GRPO reward.}
For the GRPO baseline, rewards are computed using a group of sampled responses for the same question.
Let $\mathcal{G}_x$ be the set of valid model responses sampled for question $x$ in the current rollout group.
We define the smoothed empirical model distribution as
\begin{equation}
q_{\mathrm{batch}}(y \mid x)
=
\frac{c_x(y) + \lambda}{|\mathcal{G}_x| + \lambda |\mathcal{Y}(x)|},
\end{equation}
where $c_x(y)$ is the number of sampled responses equal to option $y$, and $\lambda=0.5$ is the smoothing coefficient.

The reward for a valid response $y$ is
\begin{equation}
\begin{aligned}
r_{\mathrm{GRPO}}(x,y)
&=
\mathrm{clip}_{[-2,2]}
\Big[
\log\left(p_{\mathrm{data}}(y \mid x)+\epsilon\right)
\\
&\qquad
-
\log\left(q_{\mathrm{batch}}(y \mid x)+\epsilon\right)
\Big],
\end{aligned}
\end{equation}
where $\epsilon=10^{-6}$.
Invalid responses receive reward $-2$.

This reward is positive when an answer is under-sampled by the model relative to the human distribution and negative when it is over-sampled.
Thus, unlike the PPO reward, it uses the current rollout group to encourage closer agreement between sampled model responses and the empirical survey distribution.

\section{Supplementary WVS Held-out Evaluation}
\label{sec:appendix_wvs_holdout}

Table~\ref{tab:wvs_holdout_non_africa} reports WVS 80/20 held-out evaluation results for the non-African countries in the main study.
This supplementary evaluation uses the same within-survey split protocol as the African countries in the main evaluation and serves as a robustness check for the ISSP/WVS protocol difference.

\begin{table*}[!t]
\centering
\small
\setlength{\tabcolsep}{3pt}
\renewcommand{\arraystretch}{1.08}
\resizebox{\textwidth}{!}{
\begin{tabular}{ll|ccc|ccc|ccc|cc|c}
\toprule
{} & {} & \multicolumn{3}{c|}{\textbf{America}} & \multicolumn{3}{c|}{\textbf{Europe}} & \multicolumn{3}{c|}{\textbf{Asia}} & \multicolumn{2}{c|}{\textbf{Oceania}} & \textbf{Avg} \\
\cline{3-13}
\textbf{G} & \textbf{Model} & \textbf{CHL} & \textbf{MEX} & \textbf{USA} & \textbf{DEU} & \textbf{GBR} & \textbf{RUS} & \textbf{IND} & \textbf{JPN} & \textbf{PHL} & \textbf{AUS} & \textbf{NZL} & {} \\
\midrule
\multirow{2}{*}{$G=1$}
& Base model & 77.85 & 77.83 & 80.43 & 78.68 & 81.45 & 78.32 & 76.80 & 77.53 & 76.78 & 80.97 & 80.66 & 78.85 \\
& \textbf{ACE-Align} & \textbf{89.81} & \textbf{87.06} & \textbf{88.75} & \textbf{86.48} & \textbf{89.62} & \textbf{88.32} & \textbf{86.05} & \textbf{84.70} & \textbf{88.40} & \textbf{86.94} & \textbf{87.05} & \textbf{87.56} \scriptsize{(+8.71)} \\
\midrule
\multirow{2}{*}{$G=2$}
& Base model & 77.76 & 77.69 & 80.49 & 78.40 & 81.21 & 77.83 & 76.67 & 77.30 & 76.66 & 81.09 & 80.16 & 78.66 \\
& \textbf{ACE-Align} & \textbf{89.62} & \textbf{87.74} & \textbf{88.59} & \textbf{86.22} & \textbf{90.05} & \textbf{88.02} & \textbf{86.29} & \textbf{84.39} & \textbf{88.74} & \textbf{87.45} & \textbf{86.94} & \textbf{87.64} \scriptsize{(+8.98)} \\
\midrule
\multirow{2}{*}{$G=3$}
& Base model & 77.30 & 77.58 & 80.08 & 77.95 & 80.98 & 77.36 & 76.43 & 76.91 & 76.50 & 80.83 & 79.72 & 78.33 \\
& \textbf{ACE-Align} & \textbf{88.89} & \textbf{88.00} & \textbf{88.27} & \textbf{85.82} & \textbf{90.27} & \textbf{87.84} & \textbf{86.33} & \textbf{83.97} & \textbf{88.62} & \textbf{87.45} & \textbf{86.86} & \textbf{87.48} \scriptsize{(+9.15)} \\
\midrule
\multirow{2}{*}{$G=4$}
& Base model & 76.96 & 77.27 & 79.44 & 77.41 & 80.58 & 76.80 & 76.03 & 76.50 & 76.24 & 80.35 & 79.14 & 77.88 \\
& \textbf{ACE-Align} & \textbf{88.73} & \textbf{87.89} & \textbf{87.94} & \textbf{85.04} & \textbf{90.25} & \textbf{87.46} & \textbf{86.11} & \textbf{83.98} & \textbf{88.08} & \textbf{87.20} & \textbf{86.82} & \textbf{87.23} \scriptsize{(+9.35)} \\
\bottomrule
\end{tabular}
}
\caption{Supplementary WVS 80/20 held-out evaluation for non-African countries. The average is computed over the 11 non-African countries in the main study.}
\label{tab:wvs_holdout_non_africa}
\end{table*}
\FloatBarrier

\section{Additional Geographic Equity Results}
\label{sec:appendix_geographic_sensitivity}

\begin{figure}[H]
\centering
\includegraphics[width=\linewidth]{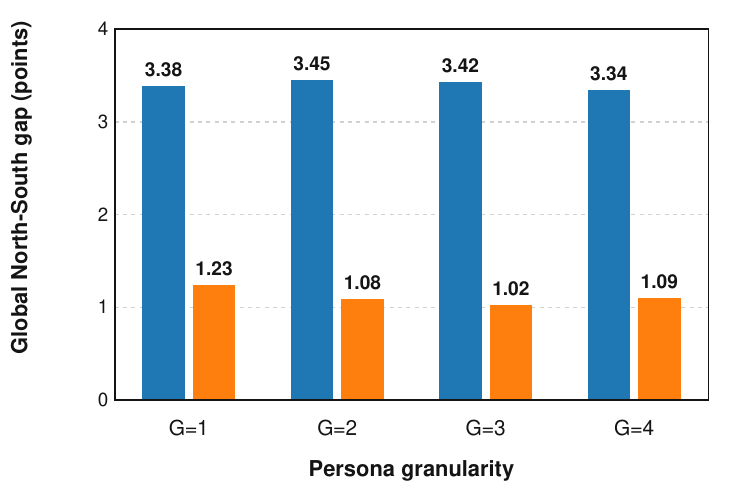}
\caption{WVS-only Global North--South alignment gaps across persona granularities under the same 80/20 held-out protocol for all 14 countries. Each bar reports mean(Global North) minus mean(Global South). Lower values indicate a smaller geographic gap.}
\label{fig:wvs_geographic_gap}
\end{figure}

Figure~\ref{fig:wvs_geographic_gap} reports the WVS-only Global North--South gaps. All 14 countries use the same WVS 80/20 held-out protocol. ACE-Align reduces the gap at every persona granularity under this within-survey comparison.

Our primary analysis follows the grouping in Figure~\ref{fig:main_results}, with Russia included in the Global South. Because Russia's contemporary classification remains debated, we conduct a sensitivity analysis that reassigns Russia to the Global North. Table~\ref{tab:geographic_sensitivity} reports the Global North--South gaps averaged across $G\in\{1,2,3,4\}$. ACE-Align reduces the gap under both groupings and within both evaluation protocols.

\begin{table}[H]
\centering
\small
\setlength{\tabcolsep}{4pt}
\renewcommand{\arraystretch}{1.08}
\resizebox{\columnwidth}{!}{
\begin{tabular}{lrrrr}
\toprule
Grouping & WVS Base & WVS ACE & ISSP Base & ISSP ACE \\
\midrule
RUS in Global South & 3.40 & \textbf{1.11} & 2.53 & \textbf{0.85} \\
RUS in Global North & 3.33 & \textbf{1.49} & 1.89 & \textbf{0.21} \\
\bottomrule
\end{tabular}
}
\caption{Sensitivity of the average Global North--South alignment gap to the geographic classification of Russia. Values are averaged across persona granularities; lower values indicate a smaller gap.}
\label{tab:geographic_sensitivity}
\end{table}

\section{Prompt Template}
\label{sec:appendix_prompt}

\begin{figure}[H]
\centering
\includegraphics[width=\linewidth]{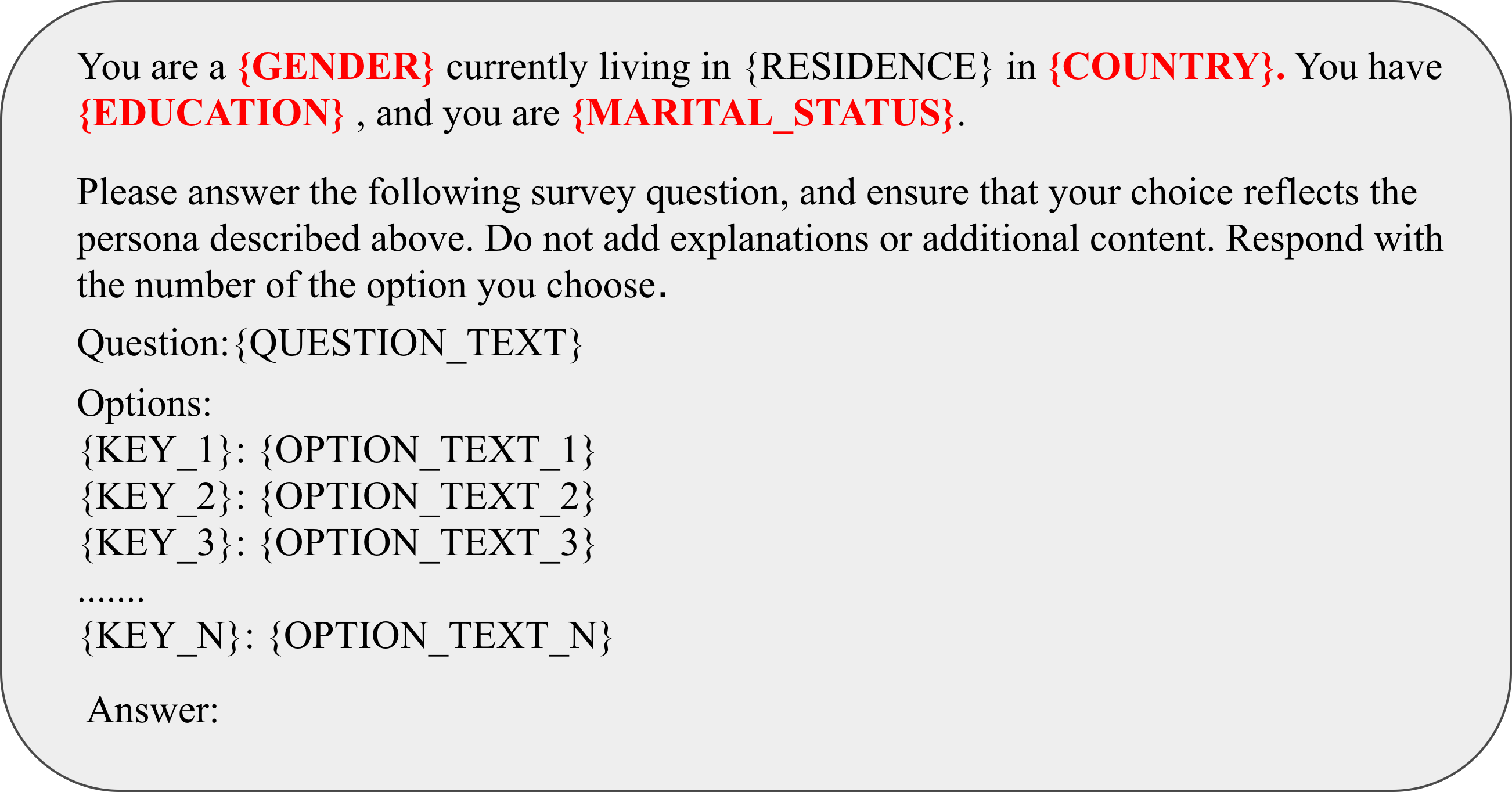}
\caption{Persona-conditioned prompt template used for cultural alignment experiments.}
\label{fig:prompt_template}
\end{figure}

Figure~\ref{fig:prompt_template} illustrates the persona prompting format used in all experiments.

\FloatBarrier

\section{Statistical Significance}
\label{sec:appendix_significance}

We assess statistical significance with country-clustered paired sign-flip tests.
For each granularity and comparison, we first compute country-level paired score differences using the same aggregation protocol and 14-country set as Table~\ref{tab:main_res}.
We compare ACE-Align with all nine baselines in Table~\ref{tab:main_res}: the base model, AdPrompt, DPO, PPO, GRPO, SimLLM, CultureSPA, Anchor-only, and Causal-only.
We then run an exact two-sided sign-flip test over the 14 country-level differences and apply Holm correction over the 36 planned comparisons in Table~\ref{tab:significance_tests}.
Confidence intervals are estimated by bootstrap resampling countries.
All mean differences and bootstrap confidence intervals are positive.
ACE-Align improves over all baselines except SimLLM and GRPO in all 14 countries at every granularity; for the stronger SimLLM and GRPO baselines, positive-country counts range from 11/14 to 13/14.

\begin{table}[H]
\centering
\scriptsize
\setlength{\tabcolsep}{2pt}
\renewcommand{\arraystretch}{1.08}
\resizebox{\columnwidth}{!}{
\begin{tabular}{llcccc}
\toprule
\textbf{G} & \textbf{Comparison} & \textbf{Mean $\Delta$} & \textbf{95\% CI} & \textbf{$p$} & \textbf{Holm $p$} \\
\midrule
$G=1$ & vs. Base model & +4.62 & [3.43, 6.11] & 0.00012 & 0.00439 \\
$G=1$ & vs. AdPrompt & +4.23 & [3.13, 5.68] & 0.00012 & 0.00439 \\
$G=1$ & vs. DPO & +4.31 & [3.34, 5.54] & 0.00012 & 0.00439 \\
$G=1$ & vs. PPO & +5.45 & [4.31, 6.90] & 0.00012 & 0.00439 \\
$G=1$ & vs. GRPO & +2.02 & [0.36, 3.36] & 0.02808 & 0.08423 \\
$G=1$ & vs. SimLLM & +1.38 & [0.17, 2.49] & 0.04407 & 0.08423 \\
$G=1$ & vs. CultureSPA & +4.26 & [3.31, 5.59] & 0.00012 & 0.00439 \\
$G=1$ & vs. Anchor-only & +5.57 & [4.54, 6.55] & 0.00012 & 0.00439 \\
$G=1$ & vs. Causal-only & +3.19 & [2.10, 4.79] & 0.00012 & 0.00439 \\
\midrule
$G=2$ & vs. Base model & +4.51 & [3.23, 6.09] & 0.00012 & 0.00439 \\
$G=2$ & vs. AdPrompt & +4.27 & [3.11, 5.74] & 0.00012 & 0.00439 \\
$G=2$ & vs. DPO & +4.21 & [3.19, 5.47] & 0.00012 & 0.00439 \\
$G=2$ & vs. PPO & +5.56 & [4.41, 7.01] & 0.00012 & 0.00439 \\
$G=2$ & vs. GRPO & +2.19 & [0.55, 3.54] & 0.01941 & 0.08118 \\
$G=2$ & vs. SimLLM & +1.42 & [0.29, 2.50] & 0.03162 & 0.08423 \\
$G=2$ & vs. CultureSPA & +4.41 & [3.47, 5.72] & 0.00012 & 0.00439 \\
$G=2$ & vs. Anchor-only & +5.79 & [4.81, 6.83] & 0.00012 & 0.00439 \\
$G=2$ & vs. Causal-only & +3.05 & [1.88, 4.70] & 0.00012 & 0.00439 \\
\midrule
$G=3$ & vs. Base model & +4.66 & [3.36, 6.23] & 0.00012 & 0.00439 \\
$G=3$ & vs. AdPrompt & +4.47 & [3.29, 5.94] & 0.00012 & 0.00439 \\
$G=3$ & vs. DPO & +4.18 & [3.16, 5.43] & 0.00012 & 0.00439 \\
$G=3$ & vs. PPO & +5.45 & [4.33, 6.89] & 0.00012 & 0.00439 \\
$G=3$ & vs. GRPO & +2.41 & [0.84, 3.68] & 0.00952 & 0.06665 \\
$G=3$ & vs. SimLLM & +1.55 & [0.46, 2.58] & 0.01624 & 0.08118 \\
$G=3$ & vs. CultureSPA & +4.61 & [3.69, 5.86] & 0.00012 & 0.00439 \\
$G=3$ & vs. Anchor-only & +5.95 & [5.06, 6.95] & 0.00012 & 0.00439 \\
$G=3$ & vs. Causal-only & +3.03 & [1.84, 4.67] & 0.00012 & 0.00439 \\
\midrule
$G=4$ & vs. Base model & +4.82 & [3.56, 6.36] & 0.00012 & 0.00439 \\
$G=4$ & vs. AdPrompt & +4.66 & [3.51, 6.07] & 0.00012 & 0.00439 \\
$G=4$ & vs. DPO & +4.12 & [3.13, 5.33] & 0.00012 & 0.00439 \\
$G=4$ & vs. PPO & +5.27 & [4.20, 6.65] & 0.00012 & 0.00439 \\
$G=4$ & vs. GRPO & +2.61 & [1.03, 3.87] & 0.00720 & 0.05762 \\
$G=4$ & vs. SimLLM & +1.64 & [0.58, 2.62] & 0.01038 & 0.06665 \\
$G=4$ & vs. CultureSPA & +4.83 & [3.92, 5.99] & 0.00012 & 0.00439 \\
$G=4$ & vs. Anchor-only & +6.09 & [5.24, 7.06] & 0.00012 & 0.00439 \\
$G=4$ & vs. Causal-only & +3.07 & [1.91, 4.63] & 0.00012 & 0.00439 \\
\bottomrule
\end{tabular}
}
\caption{Country-clustered paired significance tests for the planned ACE-Align comparisons. Mean $\Delta$ denotes the average country-level score difference in percentage points.}
\label{tab:significance_tests}
\end{table}
\FloatBarrier

\section{Placebo Test for Causal Intervention Validity}
\label{sec:appendix_placebo}

We run a placebo intervention test to examine whether controlled prompt edits induce attribute-specific distributional shifts rather than generic sensitivity to superficial prompt changes.
Using the base model in the United States setting, we compare three matched-pair interventions across all 10 survey categories and persona granularities $G\in\{1,2,3,4\}$.
The real intervention toggles \textit{Education} (college educated vs.\ not college educated), while two placebo interventions toggle semantically irrelevant prompt fields: \textit{respondent ID} (Respondent A vs.\ Respondent B) and \textit{name initial} (names starting with A vs.\ B).
For each intervention, the paired prompts differ only in the target field, and we measure the induced output-distribution shift with Earth Mover's Distance (EMD).

Across 14,826 matched pairs per intervention group, the real education intervention yields a mean $\Delta$EMD of $9.44\times 10^{-2}$, whereas the respondent-ID and name-initial placebos yield only $1.55\times 10^{-2}$ and $1.24\times 10^{-2}$, respectively.
As shown in Table~\ref{tab:placebo_test}, this 6--9$\times$ gap is consistent across all persona granularities.
Welch's $t$-tests confirm that the gap between the real intervention and each placebo is significant across all comparisons ($t>149$, $p<0.001$), with large effect sizes (Cohen's $d>1.7$).
The small residual shifts under placebo interventions provide an estimate of baseline prompt-perturbation noise, while the substantially larger education shifts support the use of controlled persona edits as meaningful attribute interventions in our operational LLM setting.

\begin{table*}[t]
\centering
\small
\setlength{\tabcolsep}{5pt}
\renewcommand{\arraystretch}{1.08}
\resizebox{\textwidth}{!}{
\begin{tabular}{lrrrrrr}
\toprule
\textbf{G} & \textbf{Real $\Delta$EMD} & \textbf{Placebo 1 $\Delta$EMD} & \textbf{Placebo 2 $\Delta$EMD} & \textbf{Real/P1} & \textbf{Real/P2} & \textbf{Cohen's $d$ vs. P1/P2} \\
\midrule
1 & 11.86 & 1.70 & 1.33 & 7.0 & 8.9 & 1.86 / 1.94 \\
2 & 10.53 & 1.74 & 1.37 & 6.1 & 7.7 & 1.79 / 1.87 \\
3 & 9.37 & 1.58 & 1.26 & 5.9 & 7.5 & 1.75 / 1.83 \\
4 & 8.42 & 1.34 & 1.10 & 6.3 & 7.6 & 1.72 / 1.79 \\
\bottomrule
\end{tabular}
}
\caption{Placebo intervention test for prompt-based causal intervention validity. $\Delta$EMD values are multiplied by 100 for readability. Placebo 1 toggles respondent ID, and Placebo 2 toggles name initial. All real-vs.-placebo comparisons have $p<0.001$.}
\label{tab:placebo_test}
\end{table*}
\FloatBarrier

\clearpage

\end{document}